\documentclass{JINST}

\usepackage{amssymb,graphicx,xspace,url}
\newcommand\tr{\mathcal{T}}
\newcommand{\ud}{\mathrm{d}}

\title{On the possiblity of using vertically pointing Central Laser Facilities to calibrate the Cherenkov Telescope Array}
\author{Markus Gaug$^{a,b}$\\
\llap{$^a$}F\'isica de les Radiacions, Departament de F\'isica, Universitat Aut\`onoma de Barcelona, 08193 Bellaterra, Spain.\\
\llap{$^b$}CERES, Universitat Aut\`onoma de Barcelona-IEEC, 08193 Bellaterra, Spain.\\
 E-mail: \email{markus.gaug@uab.cat}}

\abstract{%
A Central Laser Facility is a system composed of a laser placed at a certain distance from a light–detector array, emitting fast light pulses, 
typically in the vertical direction, with the aim to calibrate that array. During calibration runs, all detectors are pointed towards the same portion of the laser beam 
at a given altitude. Central Laser Facilities are used for various currently operating ultra-high-energy cosmic ray and imaging atmospheric Cherenkov telescope arrays. 
In view of the future Cherenkov Telescope Array, a similar device could provide a fast calibration of the whole installation at different wavelengths.  
The relative precision (i.e. each individual telescope with respect to the rest of the array is expected) to be better than 5\%, while an absolute
calibration should reach a precisions of 4--11\%, if certain design requirements are met. 
Additionally, a preciser monitoring of the sensitivity of each telescope can be made on time-scales of days to years.}

\keywords{Central Laser Facility;CTA;Gamma Ray Astronomy;Instrumentation and Methods for Astrophysics;IACT}

\begin{document}

\section{Introduction}

Central Laser Facilities (CLF) have been used widely to calibrate fluorescence detectors, like HiRes~\cite{HiRes2006}, 
AUGER~\cite{auger2006,auger2011,auger2013} and the Telescope Array~\cite{ta2009,ta2012}. 
Such facilities employ a laser to emit fast light pulses of precisely monitored power, mostly in the vertical direction, 
although the more modern systems also incorporate a steerable beam option. 
The scattered laser light 
received by the photomultipliers of the detectors resembles that from fluorescing ultra high energy cosmic ray shower tracks and is 
used to calibrate the response of the photo-detector to these. CLFs are therefore ideal calibration devices for the fluorescence detectors, and are 
routinely run several times a night.

In the case of Imaging Atmospheric Cherenkov Telescopes (IACTs)~\cite{weekes2005,buckley2008,hinton2009,holder2012,hillas2013}, 
a part of the laser path is seen as a track traveling across the focal plane of the camera, on micro-second time scales. 
IACTs consist nowadays of several telescopes and are optimized for the observation of Cherenkov light from air showers that are observed head-on
and yield light pulses with full-width-half-maxima of typically few nanoseconds, with increasing pulse lengths, as the triggerable shower impact distance 
-- and hence the camera fields-of-view -- become larger. In extreme cases, several tens of nanoseconds pulse widths may be reached~\cite{cta}. 
The air shower images are then mainly (but not exclusively) used for 
gamma-ray astronomy in the energy range from tens of GeV to hundreds of TeV. Telescope calibration by a CLF is then only useful if 
the light pulses from CLF tracks are amplified and electronically transmitted and digitized in the same, undistorted way as the 
shorter Cherenkov light pulses. Moreover, the telescopes must be able to trigger on, and buffer, the much slower moving signals through the camera.
In this case, a CLF can be used to monitor the sensitivity of each individual telescope, including mirrors and camera, 
and to cross-calibrate telescopes, or telescope types, between each other.
Finally, an absolute calibration of the whole array can be attempted. 
Contrary to the already existing CLFs, we propose to operate a CLF at multiple wavelengths, allowing for a full spectral characterization 
of each telescope. 
The calibration could be carried out during selected very clear nights -- ocurring frequently at astronomical sites -- on time-scales of about once per month.
Such a calibration scheme has the advantage to be fast and relatively cheap, as only one, or very few, devices are involved for the entire array. 
It cannot be used to equalize the gains of individual pixels of a camera, a task for which an individual light source for each telescope 
is better suited~\cite{gaug,hanna,aharonian,gaugphd,piron,rovero,puhlhofer}. 

VERITAS is the only IACT that has explored calibration of its Cherenkov telescopes with the help of a CLF~\cite{veritas2005,veritas2008}. 
A dedicated trigger and readout scheme has been developed there, which allows 
to read the signal of each pixel at different memory depths. While VERITAS has not yet published the precision of the achieved calibration, the AUGER 
collaboration has reached an absolute calibration precision better than 10\% using this technique~\cite{augerabsolute1,augerabsolute2}, 
and the Telescope Array cites 7.2\%~\cite{ta2012}. All installations make use, however, of other calibration devices and rely on their CLFs to yield redundant information. 
The AUGER experiment, moreover, uses different types of lasers to characterize the atmosphere and the fluorescence detectors. 

Given the experience of these installations, we now discuss a possible use of a CLF for the future Cherenkov Telescope Array (CTA)~\cite{cta,ctaconcept,ctamc}. 
The CTA will consist of a Southern installation, covering about 10~km$^2$ with telescopes, and a Northern one of about 1~km$^2$ extension. 
While the Southern array will contain at least three different telescope types, employing different mirror dish sizes and fields-of-view, 
the smaller Northern array is currently foreseen to consist of two types of telescopes. In both cases, 
large-size telescopes will be located in the center, surrounded by medium-size and small-size telescopes, the latter only for the Southern array.
The aim of this paper is to predict the foreseen precision with which the sensitivities of individual telescopes can be monitored and calibrated against each other, 
as well as the absolute calibration precision for the entire observatory.

\section{Geometrical considerations \label{sec:geom}}

The basic idea of a calibrating device for a telescope array is that all telescopes observe the same light source, in our case the same part of a laser beam. 
With an array of telescopes distributed over an area of 1--10~km$^2$, this is obviously impossible, however one can try to make them observe a 
very similar part of the laser beam, and to apply only small corrections for each telescope. \\
\par\noindent
In the case of a CLF, each telescope camera observes the reflected light from the laser beam, within a path length defined 
by the field-of-view (FOV) of the camera. The laser beam is then seen in the camera traveling as a stripe from the uppermost part of the camera down to the lowest part.
Each pixel along that stripe will observe a part of the laser beam corresponding to its FOV. 
The accumulated charge in an illuminated pixel reflects then the output power of the laser, the scattering and absorption of laser light in 
the atmosphere and the telescope sensitivity to light at the wavelength of the laser. If the first two parts can be controlled to a precision better 
than the known telescope sensitivities,
the CLF can ultimately serve for an absolute calibration of each telescope or the whole observatory. 



The geometry of the setup has to be chosen such that each camera 
observes roughly the same part of the laser beam.
At the same time, one should avoid observing the beam at a place where scattering of light is strongly influenced by aerosols, 
i.e. the observed part of the laser beam has to be always above the Planetary Boundary Layer, or above the Nocturnal Boundary Layer, if present. 
For precise measurements, care should also be taken not to focus the telescopes in the direction of low galactic and ecliptic latitudes in order 
to avoid strong influence of star light or zodiacal light.
Hence a minimum height of $\sim$2~km is required, depending on the precise atmospheric conditions on site. 
One could now think of a very close device, even one situated at the very center of the CTA observatory. However, in this case the closest 
telescopes will observe an almost infinite part of the laser beam, since the cotangent of the zenith angle of observation and the FOV of 
each pixel are involved. On the other hand, a very distant device, observed at a high zenith angle, makes the observed laser path 
long and extending to very high altitudes. In the end, an optimal distance has to be found which yields the smallest number of corrections for the 
different calibration steps.
\par\noindent
\begin{figure}[h!]
\centering
\includegraphics[width=0.99\linewidth]{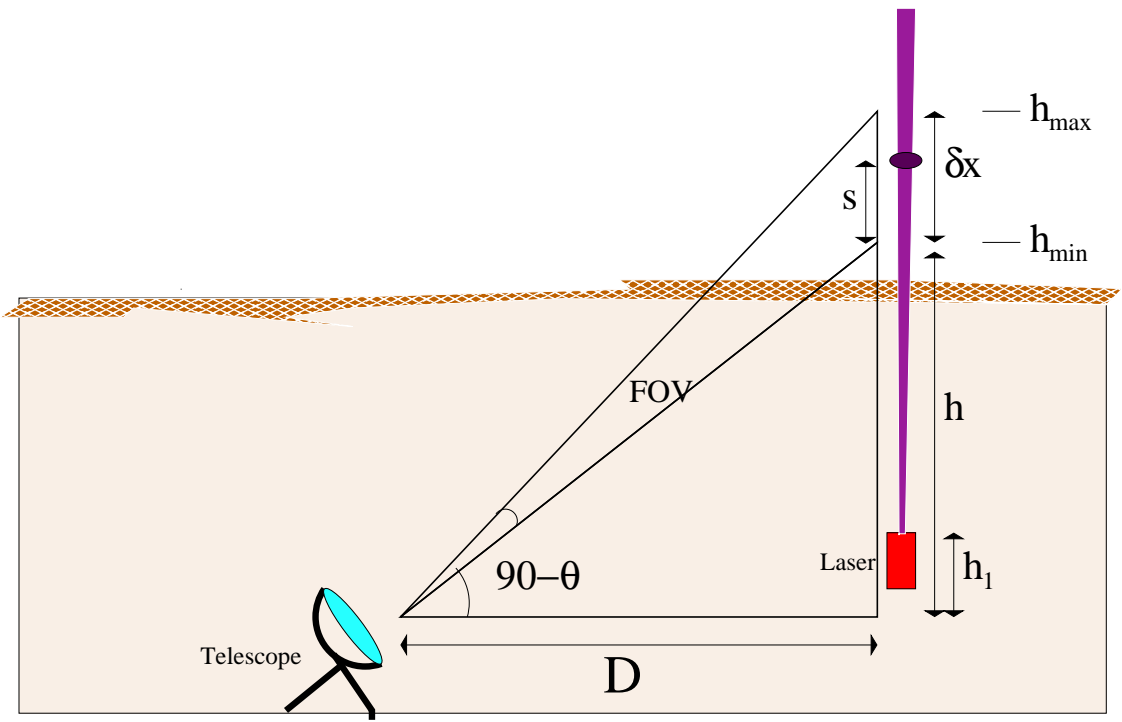}
\caption{Sketch of the introduced geometry: At the left side of the triangles, a telescope with a given FOV observes a portion of length $\delta x$ of the laser 
beam, under the zenith angle of observation $\theta$. The laser can be located at an altitude $h_1$, different from the telescope. A laser pulse is shown moving 
from $h_\mathrm{min}$ to $h_\mathrm{max}$, currently at a distance $s$ from $h_\mathrm{min}$. The brown band in the background sketches the nocturnal boundary layer.\label{fig1}}
\end{figure}
Figure~\ref{fig1} shows a sketch of the introduced geometry:  A pixel of one CTA camera, or a complete camera, sees the laser beam above a height $h$ from ground, 
under a zenith angle $\theta$. 
If the telescope is situated at a distance $D$ from the CLF laser, the pixel or camera will observe photons from a laser path length $\delta x$ in the atmosphere 
and the following relation holds:
\begin{eqnarray}
\delta x      &=& \frac{D^2+h^2}{D/\tan(\mathrm{FOV}) - h} \qquad . \label{eq:dx}
\end{eqnarray}


\begin{figure}[h!]
\centering
\includegraphics[width=0.495\linewidth]{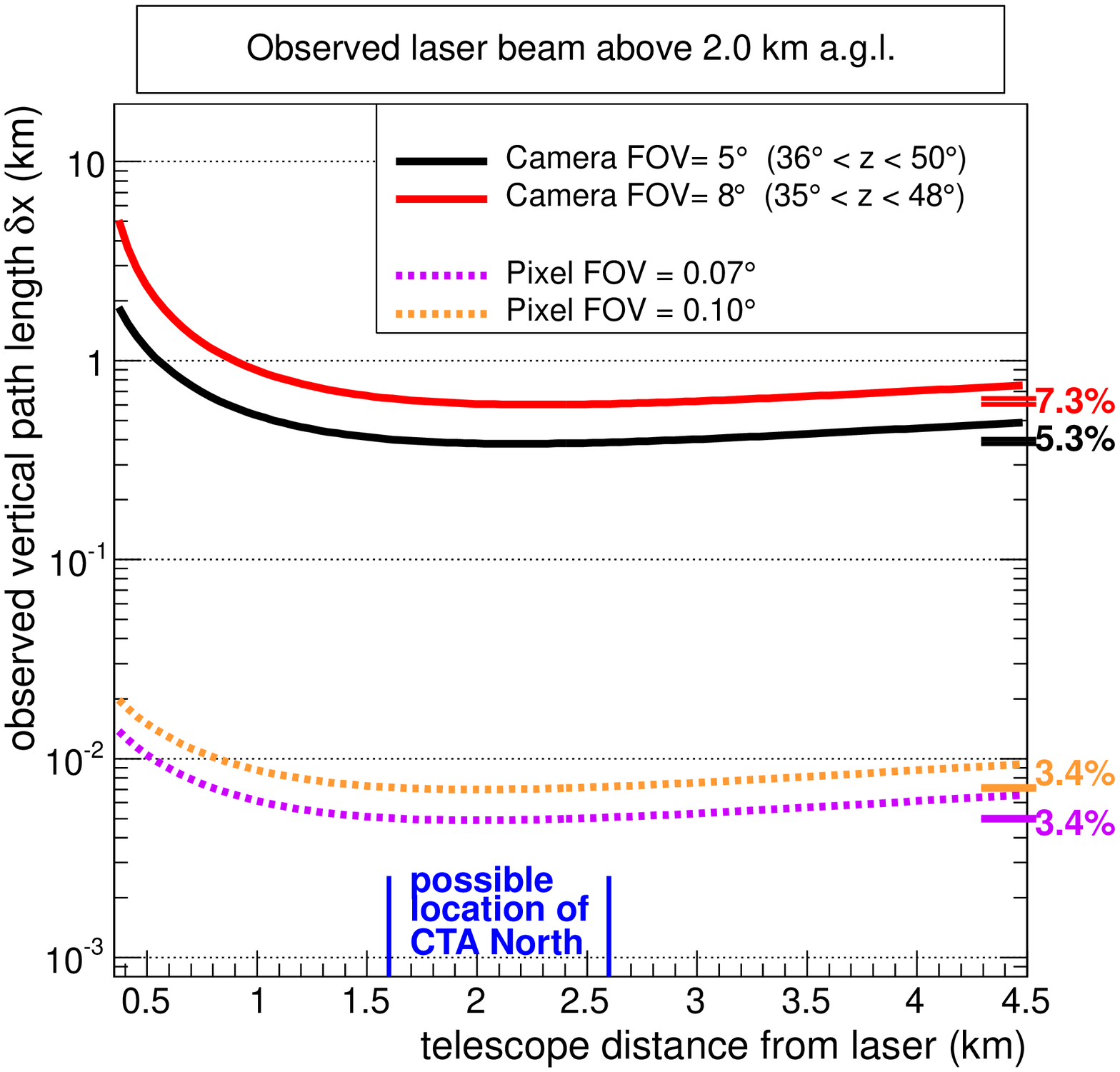}
\includegraphics[width=0.495\linewidth]{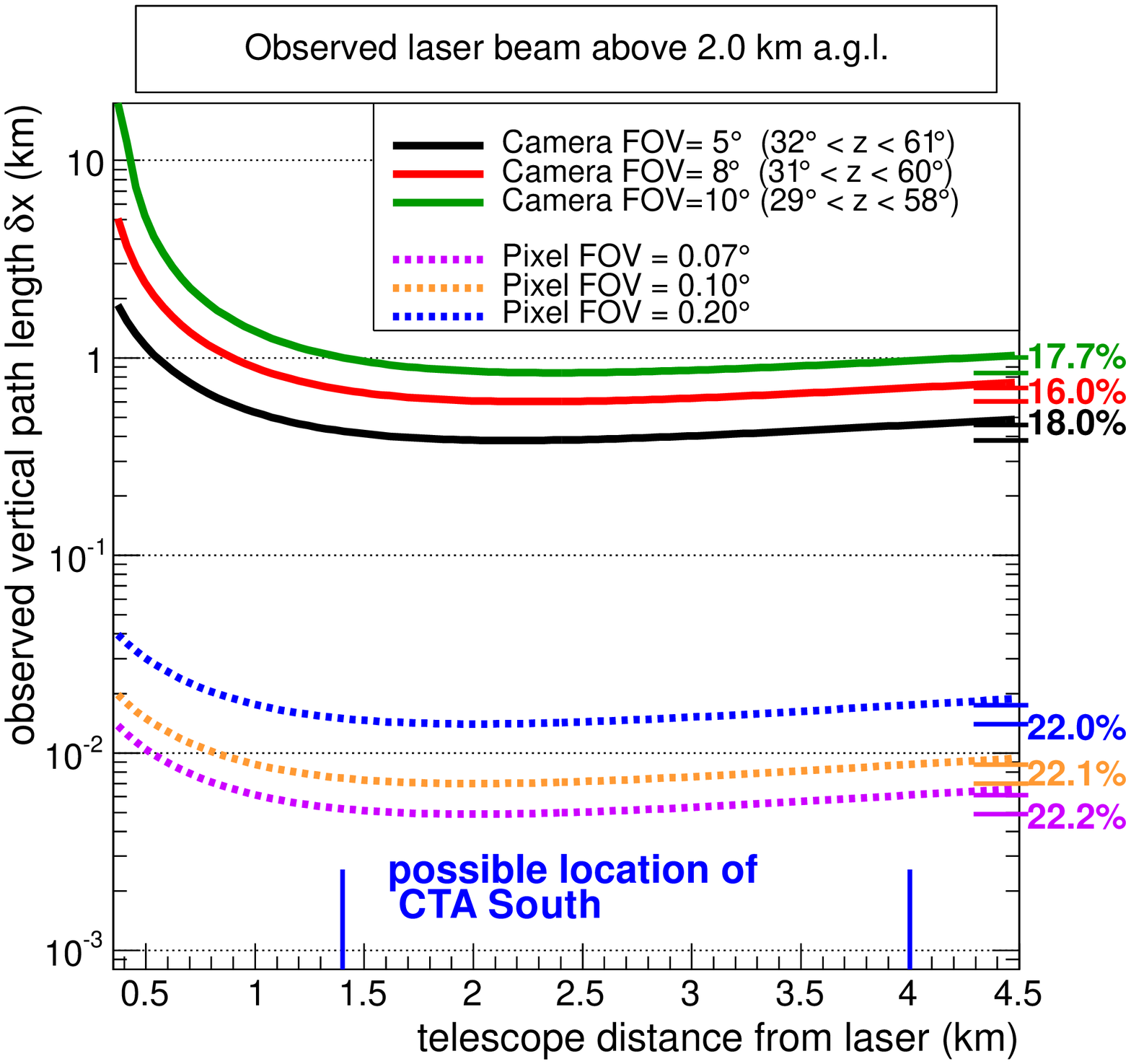}
\caption{Equation~\protect\ref{eq:dx} plotted for different FOV values for the camera (full lines) and single pixels (dotted lines), 
as a function of the telescope distance to the laser facility. An observation height of 2~km was assumed.
At the bottom in the center, the possible locations of the 
CTA (left: Northern array, right: Southern array) are suggested, 
which correspond roughly to the position of the minimum of the shown curves. 
At the right side, the relative differences in path length
are printed, between the minimum and maximum value of $\delta x$ within the suggested position of the CTA.
The legend shows also the corresponding range of zenith angles (see text for details). \label{fig2}}
\end{figure}

\noindent
Figure~\ref{fig2} shows the behavior of $\delta x$ for different discussed FOV design values~\cite{cta}, 
when plotted against the distance $D$ of a telescope with respect to the CLF laser. 
There is a broad minimum of $delta x$ found between around 1.6 and 2.6~km, which could be a possible location of the CTA-North with respect to the laser 
(Figure~\ref{fig2} left). 
The observed path lengths of the laser beam differ by less than 8\,\%, for all telescopes of a same type, even in the case of a 8$^\circ$~wide-field camera. 
The suggested solution for the position of CTA means that the telescope closest to the laser facility observes the laser track under a zenith angle of around $33^{\circ}$, 
depending slightly on the camera FOV, and the farthest telescope points to the laser beam under a zenith angle of around $50^{\circ}$. 
For the more extended Southern array, telescope positions from 1--5~km for the CLF are proposed, yielding observed laser beam length differences between 
30\% and 50\%, the latter for the case of the small telescopes with a 10$^\circ$ FOV.

\begin{figure}[h!]
\centering
\includegraphics[width=0.495\linewidth]{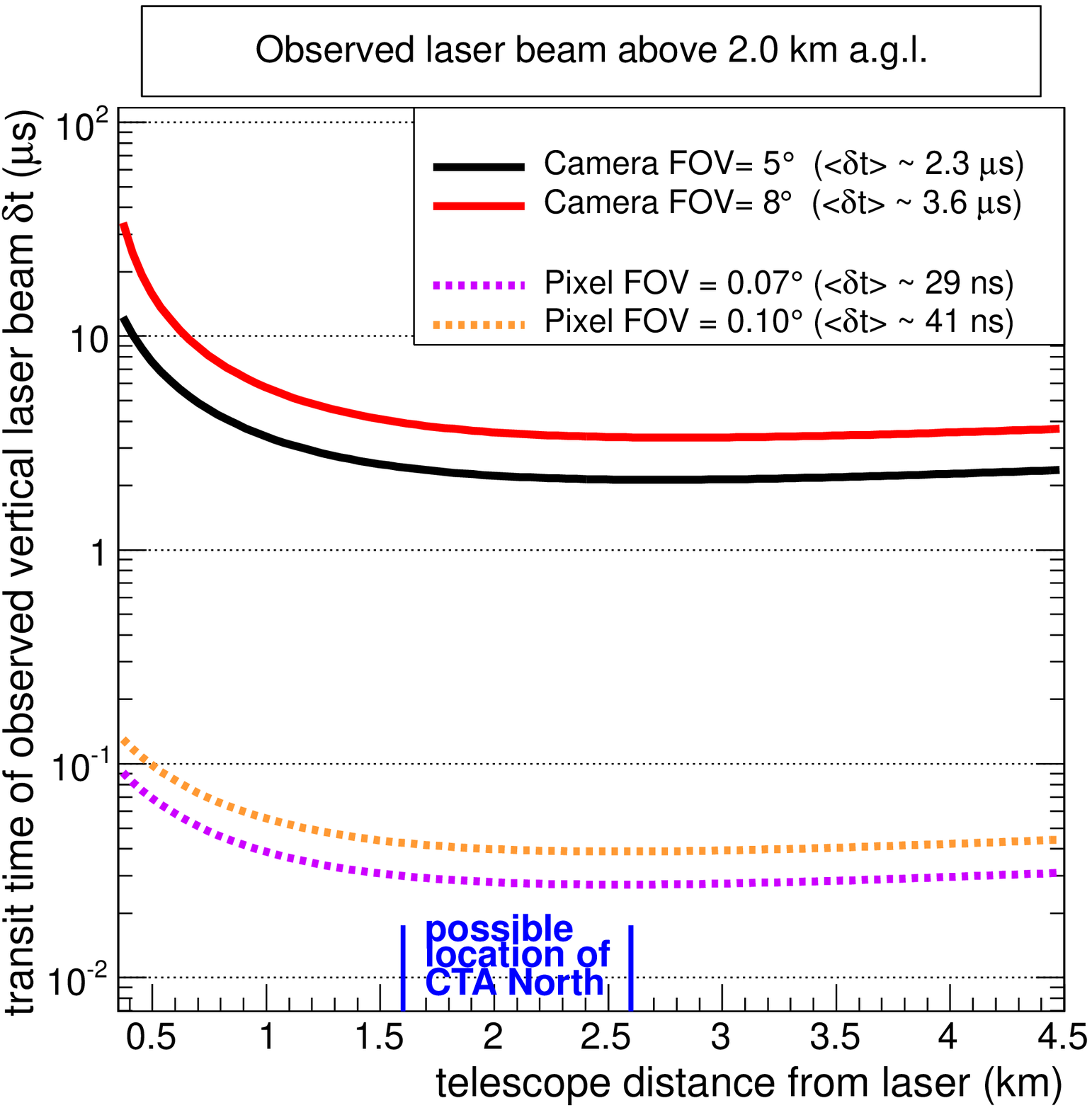}
\includegraphics[width=0.495\linewidth]{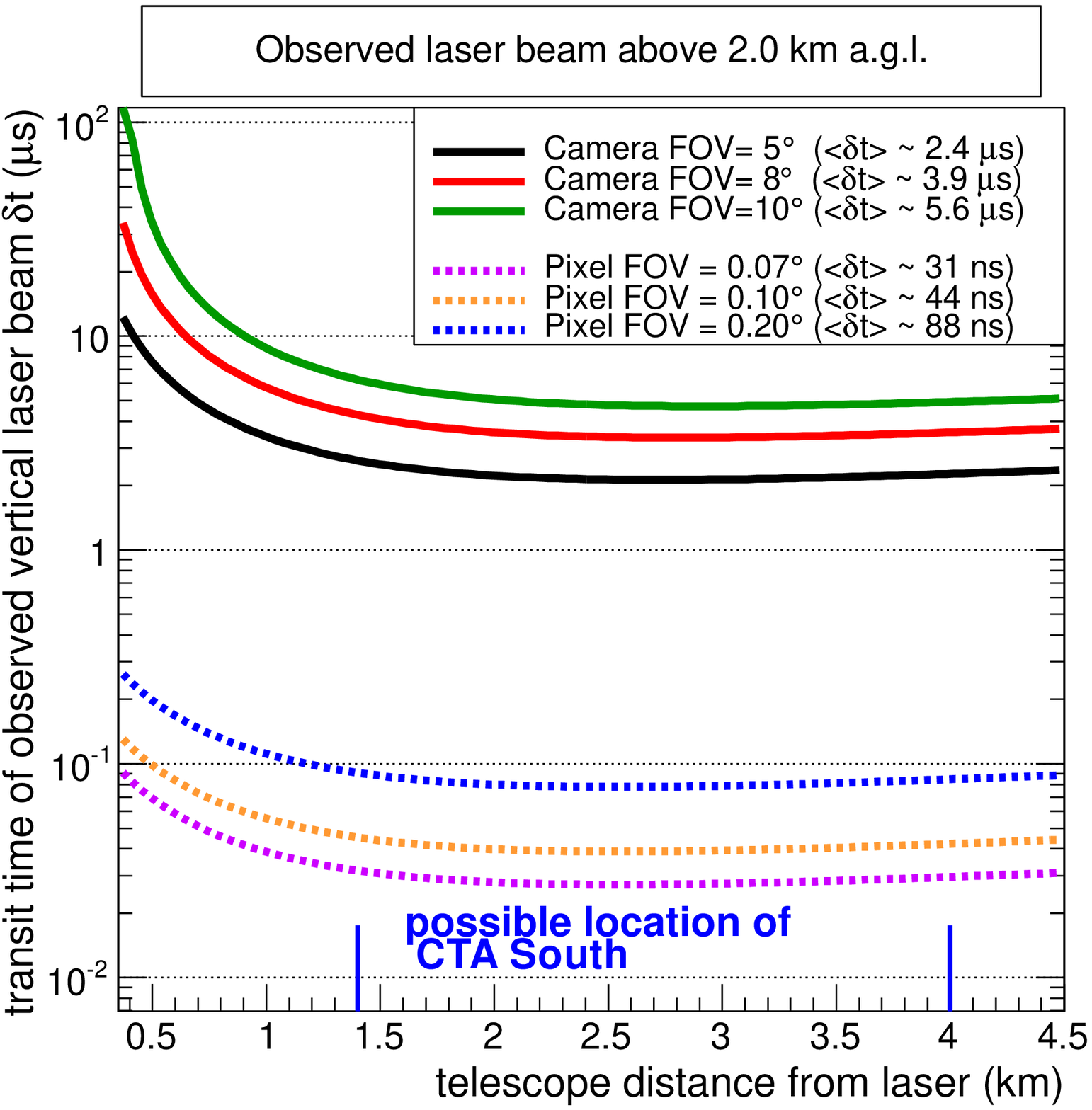}
\caption{Equation~\protect\ref{eq:dtfov} plotted for different FOV values for the camera (full lines) and single pixels (dotted lines), 
as a function of the telescope distance to the laser facility. An observation height of 2~km was assumed.
At the bottom in the center, the possible locations of the 
CTA (left: Northern array, right: Southern array) are suggested, 
which correspond roughly to the position of the minimum of the shown curves in figure~\protect\ref{fig2}.
The legend shows also the mean time delay corresponding to the suggested location of the CTA. \label{fig3}}
\end{figure}

\par\noindent
The individual pixels will see light pulses with a duration $\delta t$, larger than in the case of Cherenkov light, namely:
\begin{equation}
\delta t   = \frac{1}{c_\textrm{air}} \cdot \left( \sqrt{(h+\delta x)^2+D^2} + \delta x - \sqrt{h^2+D^2} \right) \quad , \label{eq:dt}
\end{equation}
\noindent
where $c_\textrm{air}$ is the speed of light in air. Equation~\ref{eq:dt} can also be expressed as a function of the pixel FOV:
\begin{equation}
\delta t = \frac{\sqrt{h^2+D^2}}{c_\textrm{air} \cdot (D/\tan(\mathrm{FOV})-h) } \cdot \left( \sqrt{h^2+D^2} + D\cdot \frac{1-\cos(\mathrm{FOV})}{\sin(\mathrm{FOV})}  + h \right) \quad .
\label{eq:dtfov}
\end{equation}
\noindent
Figure~\ref{fig3} shows the typical transit times of the light pulses for the currently used design FOV values for the 
 different telescope types of the CTA. One can see that the times are of the order of tens of nano-seconds for the individual pixels 
and several micro-seconds for the entire camera. These numbers can only be reduced, if the height of the observed laser path $h$ is lowered, at the 
cost of a bigger contribution of aerosols to the scattering of light into the camera.

\section{Approximating the precision of a CLF}

If we assume that the laser light is sufficiently collimated, such that the observed beam width is always smaller than the FOV of a single pixel, 
we can define the amount of light received by an individual pixel:
\begin{equation}
I_\mathrm{pix} = I_0 \cdot \frac{A_\mathrm{eff,tel}}{\sin\theta}  \int_{h_\mathrm{min}}^{h_\mathrm{max}} \beta (s) \cdot (1+f) \cdot \frac{1}{D^2+s^2} \cdot \tr_\mathrm{mol}(s) \cdot \tr_\mathrm{aer}(s)\ \ud s + F_\mathrm{bgr} \quad , \label{eq:Ipix}
\end{equation}
\noindent
where $I_0$ is the light intensity at the laser source, $A_\mathrm{eff,tel}$ the effective collection area of the telescope, which includes all 
reflection and photo-detection efficiencies and light losses, and is the term to be calibrated. 
Inside the integral of equation~\ref{eq:Ipix}, $\beta (s) = \sigma_\mathrm{tot}(s) \cdot N(s) $ denotes the total volume scattering cross-section at height $s$, 
containing the contribution from molecular (Rayleigh) and aerosol (Mie) elastic and in-elastic scattering into a solid angle $\Omega(s)=A_\mathrm{eff,tel}/(D^2+s^2)$. 
The term $(1+f)$ accounts for higher order corrections 
due to multiple scatterings and $\tr_\mathrm{mol}$ and $\tr_\mathrm{aer}$ are the molecular and aerosol transmission coefficients from the laser to point $s$ and from 
there to the telescope mirror. The integration limits $h_\mathrm{min}$ and $h_\mathrm{max}$ can 
be expressed as $h_\mathrm{min} = D\cdot\cot(\theta+\mathrm{FOV}/2)$ and $h_\mathrm{max} = D\cdot\cot(\theta-\mathrm{FOV}/2)$ for a central pixel or the whole camera, otherwise 
the position of the pixel inside the camera has to be taken into account: 
$h_\mathrm{min} =D\cdot\cot(\theta+\textit{pos}_y+\mathrm{FOV}/2)$ and $h_{max} =D\cdot\cot(\theta+\textit{pos}_y-\mathrm{FOV}/2)$, where $\textit{pos}_y$ denotes the 
$y$-coordinate of the pixel position in the camera, expressed in radians. \\

\noindent
The last term $F_{bgr}$ denotes ambient light which sneaks into the detector and has to be approximated by the light 
of the night sky, accumulated during the typical integration windows plus possible contributions from star light. 
Equation~\ref{eq:Ipix} assumes that the observed path length $\delta x$
is much larger than the laser pulse diameter $c_\mathrm{air} \cdot t_\mathrm{pulse}$, which can be easily achieved with current pulsed nitrogen or Nd:YAG lasers showing 
typical pulse duration values of $t_\mathrm{pulse} \sim 2-3$~ns. Further, the pulse is assumed to be \textrm{un-polarized}, and hence scattering uniform in azimuth.


%
%

%
%
%
%
%



%
%
\noindent
The transmittances $\tr$ can be split in three terms: from the laser to height $h_\mathrm{min}$ ($\tr_\mathrm{prev}$), from there to the scattering height $\tr_s(s)$ and 
finally from the scattering height to the photo-detector $\tr_\mathrm{post}$:
\begin{equation}
\tr(s) = \tr_\mathrm{prev} \cdot \tr_s(s) \cdot \tr_\mathrm{post}(s) \quad . \label{eq:transm}
\end{equation}
\noindent
While the first term $\tr_\mathrm{prev}$ is equal for all telescopes and pixels, the second term $\tr_s$ depends on the position of a pixel inside the camera, and the 
third term $\tr_\mathrm{post}$ depends on both the location of the telescope and the position of the camera pixel.  
The atmospheric transmission before the scattering will depend on the molecular and aerosol contents of the atmosphere above the laser: 
\begin{eqnarray}
\tr_\mathrm{prev} &=& \exp \big( - \int_{h_1}^{h_\mathrm{min}} \kappa_\mathrm{mol}(s) + \kappa_\mathrm{aer}(s)\ \,\ud s \big)\nonumber\\
{} &=& \exp \big[ - \tau_\mathrm{mol+aer}(h_\mathrm{min}-h_1) \big] \qquad , \label{eq:Tprev} 
\end{eqnarray}
\noindent
where $\kappa_\mathrm{mol}$ and $\kappa_\mathrm{aer}$ are the molecular and aerosol extinction coefficients, respectively and $\tau_\mathrm{mol+aer}$ 
the vertical optical depths, respectively.\\
\par\noindent
Assuming that between height $h_\mathrm{min}$ and $h_\mathrm{max}$, the molecular and aerosol contents do not vary significantly -- since we have previously requested that 
the entire observed laser path lies above the nocturnal boundary layer -- $\tr_s(s)$ can be simplified:
\begin{eqnarray}
\tr_s(s)           &=& \exp \big( - \int_{h_\mathrm{min}}^{s} \kappa_\mathrm{mol}(s') + \kappa_\mathrm{aer}(s') \ \,\ud s' \big) \\
\tr_s(s)           &\simeq& \exp \big( - [ \kappa_\mathrm{mol}(D\cdot\cot\theta) + \kappa_\mathrm{aer}(D\cdot\cot\theta) ] \cdot \delta x \big) \qquad . \label{eq:Tr} 
\end{eqnarray}
\noindent
Finally, $\tr_\mathrm{post}(s)$ can be written as:
\begin{eqnarray}
\tr_\mathrm{post}(s) &=& \exp \big( - \int_0^{h_\mathrm{min}+s} (\kappa_\mathrm{mol}(s') + \kappa_\mathrm{aer}(s'))/\cos\theta \,\ud s' \big) \label{eq:alphapost} \\
&\simeq& \exp \big[ - \tau_\mathrm{mol+aer}(D\cdot\cot\theta)  / \cos\theta \big] \quad . \label{eq:alphapost:simp} 
\end{eqnarray}
%
%
\noindent
In principle, equation~\ref{eq:Ipix} requires an additional integration over the zenith-angle dependent scattering cross-section, making the scattered light 
illuminate the telescope mirror under different angles at a same altitude $s$. However, considering the current design parameters, even for the large-size telescope, 
the committed error would be always well below 1\%, if Rayleigh scattering is considered. Both equations~\ref{eq:Tr} and~\ref{eq:alphapost:simp} are 
similar simplifications which introduce a maximal error of the same size. 

Finally, we consider the volume scattering cross-section $\beta_\mathrm{tot} (s)$, which contains contributions from molecular and aerosol scattering, both elastic and 
inelastic components. The inelastic component is usually several orders of magnitude 
smaller than the elasticly scattered signal and is further considered a tiny contribution to the systematic 
uncertainty of the calculated scattered light signal. Also the multiply scattered component is only relevant for 
turbid atmospheres and not considered further since a CLF would be operated only in very clear nights, hence $(1+f) \simeq 1$~(see e.g.~\cite{kovalev}). 

Equation~\ref{eq:Ipix} can be simplified, if the atmospheric parameters ($\kappa_\mathrm{mol}$,$\kappa_\mathrm{aer}$) do not change significantly over the observed 
path length $d\delta x$, especially the second integral equation~\ref{eq:Tr}: $(1-\exp(-\kappa(D\cdot\cot\theta)\cdot \delta x)) \leq 0.0005$, in order to keep the convolution error small enough. 
This requirement should be usually no problem in case of clear night shots, where $\kappa_\mathrm{aer} \lesssim 2\cdot 10^{-5} \mathrm{m}^{-1}$ at altitudes 
between 2 and 4~km~\cite{auger2013} and $\delta x \lesssim 20$~m. The molecular scattering part $\kappa_\mathrm{mol}$ changes by maximally 0.1\%.
The error made by using equation~\ref{eq:alphapost:simp} instead of the convolution of equation~\ref{eq:alphapost} inside equation~\ref{eq:Ipix} can be estimated 
to amount to always less than one percent for the worse case scenario (largest pixel FOV).  

The following simplified version of equation~\ref{eq:Ipix} therefore applies, if clear nights are chosen:
\begin{eqnarray}
I_\mathrm{pix} &\approx& I_0 \cdot \frac{A_\mathrm{eff,tel}}{\sin\theta} \cdot \tr (D\cdot \cot\theta) \nonumber \\
&& \cdot  \int_{h_\mathrm{min}}^{h_\mathrm{max}} \left( \beta_\mathrm{mol}(s(\theta))+\beta_\mathrm{aer}(s(\theta))\right) \cdot \frac{1}{D^2+s^2} \ \ud s + F_\mathrm{bgr} \qquad . \label{eq:Ipix:simp1}
\end{eqnarray}


%
%
%
 
\subsection{Rayleigh scattering}

The volume Rayleigh scattering cross section for unpolarized light is (see e.g.~\cite{bucholtz,mccartney}):

\begin{eqnarray}
\beta_\mathrm{mol} (\theta,\lambda,s)
           &=& \frac{6 \pi^2 \cdot (n(s)^2-1)^2}{N_s \cdot \lambda^4 \cdot (n(s)^2+2)^2} 
           \cdot \big( \frac{6+3\rho(s)}{6-7\rho(s)} \big) \cdot \frac{P(s)}{P_s}\cdot \frac{T_s}{T(s)} \nonumber\\
           &\cdot& \frac{3}{4}\cdot\big(\frac{2+2\rho(s)}{2+\rho(s)}\big) \cdot \big( 1 + \frac{1-\rho(s)}{1+\rho(s)}\cos^2\theta \big) 
           \qquad , \label{eq:Rayleigh} 
\end{eqnarray}
\noindent
where the first line shows the original cross-section formula, including the \textit{King correction}, due to the depolarization $\rho$ of the air molecules, and 
the correction for the different air densities at height $s$, measured through temperature $T$ and pressure $P$. 
The second line is the \textit{Chandrasekhar corrected} phase function~\cite{Chandrasekhar,mccartney}. 
Further, $n(s)$ is the refraction index of air and $N_s$ the number density of molecules per unit volume, 
at standard conditions ($N_s =2.5469\cdot 10^{25}~\mathrm{m}^{-3}$~\cite{bodhaine} at $T_s = 288.15$~K and $P_s = 101.325$~kPa). 
According to~\cite{tomasi}, both $n$ and $\rho$ depend slightly on the wavelength of light, atmospheric pressure, temperature, 
relative humidity and the concentration of CO$_2$. Assuming dry air and
standard values $P_s$ and $T_s$, 
we obtain for the Nitrogen laser at 337~nm 
$(n_s-1) = 2.871\cdot 10^{-4}$ and for the Nd:YAG laser at 355~nm $(n_s-1) = 2.855\cdot 10^{-4}$. 
The depolarization coefficients $\rho = 0.0311$ and $\rho = 0.0306$ can be used for typical atmospheric conditions at 4.5~km a.s.l. 
for the two wavelengths, respectively.


Using these numbers and the relation $9\cdot(n^2-1)^2/(n+2)^2 \approx 4\cdot(n-1)^2$, the volume scattering cross section can be written as:
%
%
\begin{equation}
\beta_\mathrm{mol} (\theta,\lambda,s) \approx 8.29 \cdot 10^{11}~ \frac{(n_s(\lambda)-1)^2}{(\lambda / \mathrm{nm})^4} \cdot \big(1 + 0.940\cdot\cos^2\theta \big) \cdot \frac{P(s)}{P_s}\cdot \frac{T_s}{T(s)}~\mathrm{m^{-1}\,Sr^{-1}} ~ .  \label{eq:Rayleigh1}
\end{equation}
The relation is precise to at least 0.5\%, with the main uncertainty stemming from the unknown water vapor content~\cite{tomasi}.
%
%

\subsection{Aerosol scattering}

The volume Mie scattering cross section for aerosols is frequently parameterized by the \textit{Henyey-Greenstein} formula~\cite{henyey}:

\begin{equation}
\beta_\mathrm{aer}(\theta,\lambda,s) = \beta_\mathrm{aer}(\lambda,s) \cdot \frac{1-g^2}{4\pi} \cdot 
\bigg(\frac{1}{(1+g^2-2g\cos\theta)^{3/2}} + f \frac{3 \cos^2\theta -1}{2\cdot (1+g^2)^{3/2}} \bigg) \quad,
\end{equation}

where $g$ varies between 0 and 1 and represents the mean value of
$\cos(\theta)$. The parameter $f$ characterizes the strength of the second component to the backward scattering peak. 
Typical values for clear atmospheres and desertic environments are $g \approx 0.6, f \approx 0.4$~\cite{louedec}, 
however precise aerosol scattering calculations require enhancements of the traditional Mie formalism
and can become very complex, especially in the range of scattering angles from 120 to 180 degrees~\cite{dubovik}.
It is therefore important to keep the overall aerosol contribution to scattering, expressed by the term $\beta_\mathrm{aer}(\lambda,s)$, 
as small as possible, by choosing sufficiently high scattering altitudes and clear nights. In this case, 
$\beta_\mathrm{aer}(\lambda=355\,\mathrm{nm},s=2000\,\mathrm{m}) < 2\cdot 10^{-6}$~m$^{-1}$ can be assumed, at least from experience at the AUGER 
site~\cite{auger2013}\,\footnote{Desertic aerosol models~\protect\cite{longtin}
or the older Elterman model~\protect\cite{Elterman:64} yield values of about an order of magntitude higher. 
Some African candidate sites for the CTA, known to be affected by dust intrusion, may require further detailed investigation whether the limit can be met.}
%
%
%
Inserting these numbers, we obtain an approximate expression for the aerosol volume scattering 
cross section:
%
%
\begin{equation}
\beta_\mathrm{aer}(\theta) \lesssim 6\cdot 10^{-8} \cdot \bigg( (1-0.88\cos\theta)^{-1.5} + 0.2 \cdot (3\cos^2\theta-1)\bigg)~\mathrm{m^{-1}} \quad.
\label{eq:Aerosols}
\end{equation}

\begin{figure}[h!]
\centering
\includegraphics[width=0.95\linewidth]{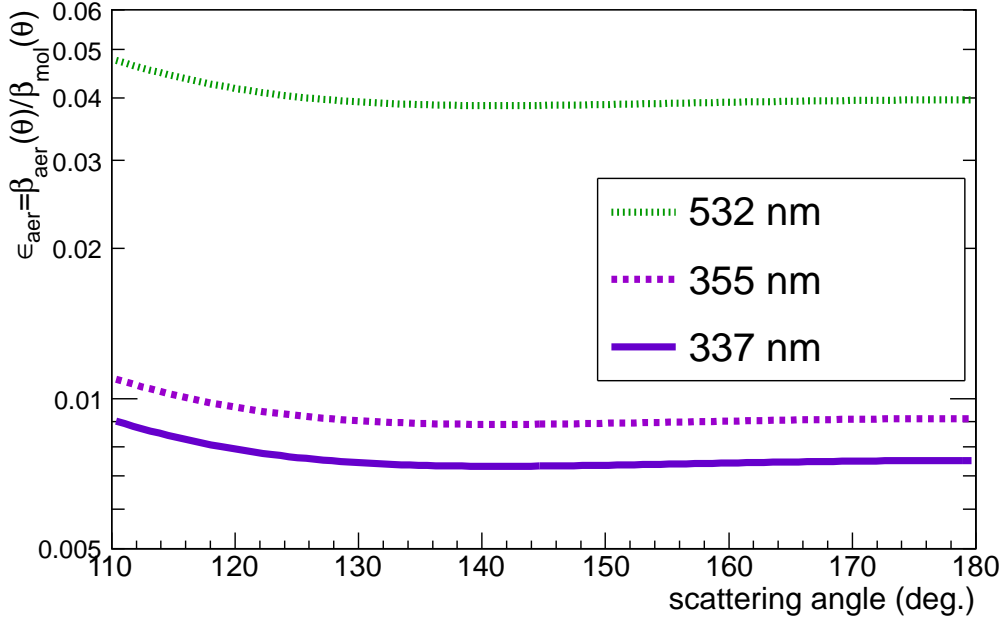}
\caption{The aerosol scattering coefficient (equation~\protect\ref{eq:Aerosols}) divided by the molecular scattering coefficient (equation~\protect\ref{eq:Rayleigh1})
 for typical conditions at 4.5~km altitude a.s.l., plotted for different scattering angles. \label{fig:epsaer}}
\end{figure}

\noindent
After inserting equations~\ref{eq:Rayleigh1} and~\ref{eq:Aerosols} into equation~\ref{eq:Ipix:simp1}, we obtain:
\begin{eqnarray}
I_\mathrm{pix} &\approx& 8.29 \cdot 10^{11} \cdot  I_0 \cdot \frac{(A_\mathrm{eff,tel}/\mathrm{m^2})}{\sin\theta} 
\cdot \frac{(n_s(\lambda)-1)^2}{(\lambda / \mathrm{nm})^4} \cdot \tr (h) \nonumber \\
&& \cdot  \int_{D\cdot\cot(\theta+\mathrm{FOV}/2)}^{D\cdot\cot(\theta-\mathrm{FOV}/2)}  \frac{1}{D^2+s^2} \cdot (1 + 0.940 \cdot \frac{s^2}{(D^2+s^2)^2}) \cdot \nonumber\\
&& \cdot  \frac{P(s)}{P_s}\cdot \frac{T_s}{T(s)} \cdot( 1 + \epsilon_\mathrm{aer}(s)) \ \ud s \nonumber \\
&& {} + F_\mathrm{bgr} \quad , \label{eq:Ipix:simp2}
\end{eqnarray}
\noindent
where $\epsilon_\mathrm{aer}$ represents the relative contribution of aerosols to the scattering process, i.e. $\epsilon_\mathrm{aer} = \beta_\mathrm{aer}(s) / \beta_\mathrm{mol}(s)$.  
As shown in fig.~\ref{fig:epsaer}, $\epsilon_\mathrm{aer} < 0.01$ in the scattering angle range of interest here, if UV laser wavelengths are used.

As the last two fractions represent the air density variations over the illuminated laser light path, and assuming the simplest model of an exponential 
decay of the density with a scale height of $H_0$ $\approx$ 9.3~km~\cite{mccartney}, one can replace the pressure and temperature corrections inside the integral by their 
mean values and make only a correction for the exponential behavior. The resulting error is always smaller than 1\%, even for the case of a 10$^{\,\circ}$~FOV camera and 
integrating the signal over the entire camera, hence: 
\begin{eqnarray}
I_\mathrm{pix} &\approx& 8.29 \cdot 10^{11} \cdot  I_0 \cdot \frac{(A_\mathrm{eff,tel}/\mathrm{m^2})\cdot \mathrm{FOV}}{(D/\mathrm{m}) \cdot \sin\theta} 
                        \cdot \frac{(n_s(\lambda)-1)^2}{(\lambda / \mathrm{nm})^4} \nonumber \\
&& \cdot \frac{P(h)}{P_s}\cdot \frac{T_s}{T(h)} \cdot \tr (h) \cdot \tr (D\cdot \cot \theta) 
  \cdot \left( 1+ 0.940 \cdot \cos^2\theta \right) \nonumber\\
&&  + F_\mathrm{bgr} + O(\epsilon_\mathrm{aer}) + O(1\%) \quad . \label{eq:Ipix:simp3}
\end{eqnarray}


%

%
\noindent
The term $(n_s(\lambda)-1)^2/\lambda^4$ in  equation~\ref{eq:Ipix:simp3} yields for the laser wavelengths 337~nm and 355~nm the values 
 $6.39\cdot 10^{-18}~\mathrm{nm}^{-4}$ and $5.13\cdot 10^{-18}~\mathrm{nm}^{-4}$, respectively, for a standard atmosphere. 
The correction factor $P(h)/P_s\cdot T_s/T(h)$ for typical altitudes of the proposed sites for the CTA array (between 1.6 and 3.5~km a.s.l.),
and assuming that the laser light is observed at 2~km height, ranges from 0.56 to 0.69. 
%
Atmospheric transmissions $\tr(D\cdot \cot \theta)$ can be estimated to range between 0.65 and 0.82 for the molecular part, and 
always greater than 0.95 for the aerosol part and clear nights\,\footnote{This number should be taken with some caution for some 
Africa candidate sites for the CTA: desertic aerosol models~\protect\cite{longtin} or the older Elterman model~\protect\cite{Elterman:64} 
yield values of about 0.85 for that case.}


Table~\ref{tab:tellist} shows predicted values of $I_\mathrm{pix}/I_0$, 
i.e. the part of the laser pulse energy which is captured by a single pixel, 
 for several currently discussed telescope realizations of the CTA and a Nitrogen laser wavelength of 337~nm.

%
%
%
\begin{table}
\centering
\begin{tabular}{lccc}
\hline
    &  $A$       & FOV$_\mathrm{pix}$ &   {\small fraction scattered light} \\
    &  (m$^2$)   &  (mrad)           &   {\small  into pixel (at 337~nm) }\\
    &            &                   &   {\small  ($\cdot 10^{-10}$)      }\\[0.27cm]
\hline
Large-Size-Telescope  &  387 & 1.8  & $(15-23)$\\
\cite{teshima2013}   &      &      &          \\
\hline
Medium-Size-Telescope &  104 & 3.1 & $(3-18)$\\
\cite{ctaconcept}    &      &     & \\
\hline
Mid-Size-Schwarzschild-Couder Tel. &  50 & 1.2 & $(0.6-2.6)$ \\
\cite{vasiliev2013}                    &     &     & \\
\hline
Small-Size-Telescope-2Mirror  & 6.0  & 3.0 & $(0.14-14)$\\
\cite{pareschi2013}    &      &     &\\
\hline
Small-Size-Telescope-2Mirror  & 8.2  & 2.6 & $(0.16-16)$\\
\cite{zech2013}        &     &      & \\
\hline
Small-Size-Telescope-1Mirror  & 7.6  & 4.4 & $(0.26-2.6)$\\
\cite{moderski2013}    &      &     & \\
\hline
\end{tabular}
\caption{\label{tab:tellist}List of possible telescope characteristics for CTA. The ``super-configuration'' of \cite{ctaconcept} has been used, 
and an overall mirror reflectivity of 90\% assumed.} 
\end{table}

%
%
Every 100~$\mu$J of laser pulse energy translates thus into about $550$ to $90,000$ photo-electrons per pixel, assuming typical values for the 
mirror reflectivity and photon collection efficiency. Using appropriate tuning of the laser power, or transmission filters in the 
optical bench after the laser output, a reasonable amount of calibration light should hence be achieved without problems using commercial 
low-power nitrogen or Nd:YAG lasers. However, the more than factor 40 difference in photon flux per pixel between the closest and the farthest medium-size-telescope
to the CLF is of concern, and can be reduced only if several CLFs are installed around the telescope array. Even more problematic is the light flux 
received by the closest medium-size-telescope, compared with the one from the farthest small-size-telescope  amounting to about a factor 100. This would 
mean cross-calibrating the telescopes with a very different signal amplitude, where possible non-linearity effects cannot be disentangled anymore from 
differences in mirror reflectivity and pixel response.

\begin{figure}[h!]
\centering
\includegraphics[width=0.95\linewidth]{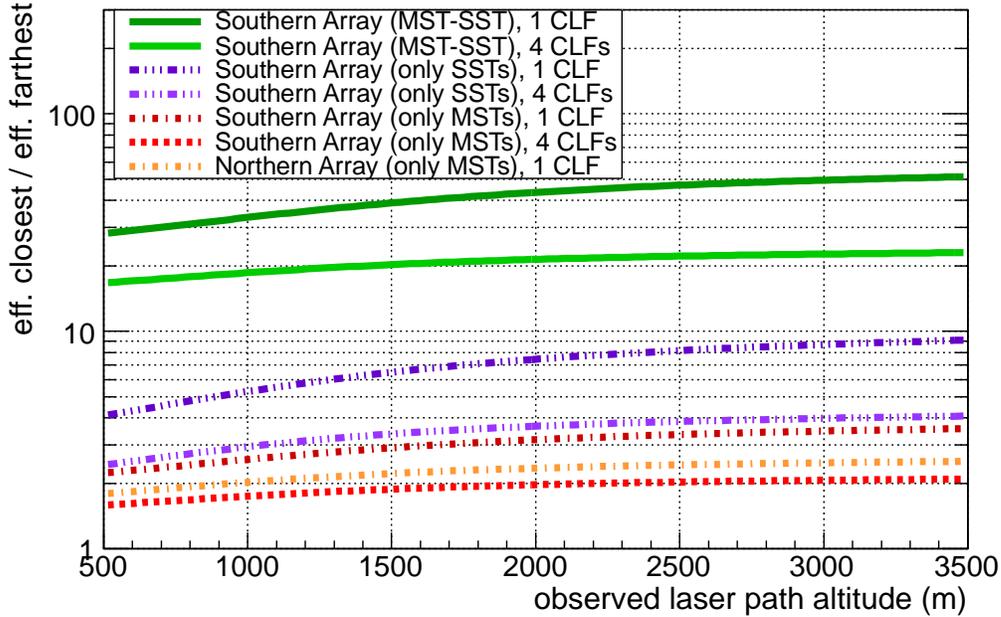}
\caption{The scattering efficiency (equation~\protect\ref{eq:Ipix:simp3}) from the closest medium-size-telescope, divided by the efficiency of the farthest telescope, for various 
observation altitudes and telescope types: The full green lines correspond to the closest medium-size-telescope, compared to the farthest small-size-telescope, 
the dashed-triple-dotted violet lines to the small-size-telescope only, and the dashed-dotted red lines to medium-size-telescopes only. 
The light green and light violet lines correspond to the same situation for the 
case of 4 central laser facilities, installed at each side of the array. 
The orange dotted line correspond to the Northern array, considering only medium-size-telescopes. \label{fig:epsdiff}}
\end{figure}

Figure~\ref{fig:epsdiff} shows that the difference in the received amount of light depends on the observation height of the laser pulse and can be reduced 
by roughly 50\% if the laser pulse is observed at 1~km height, instead of 2~km in the case of the Southern array. In the case of the smaller Northern array, 
which in its current design does not host small-size-telescopes, only differences up to a factor 2 are expected. 

\begin{figure}[h!]
\centering
\includegraphics[width=0.85\linewidth]{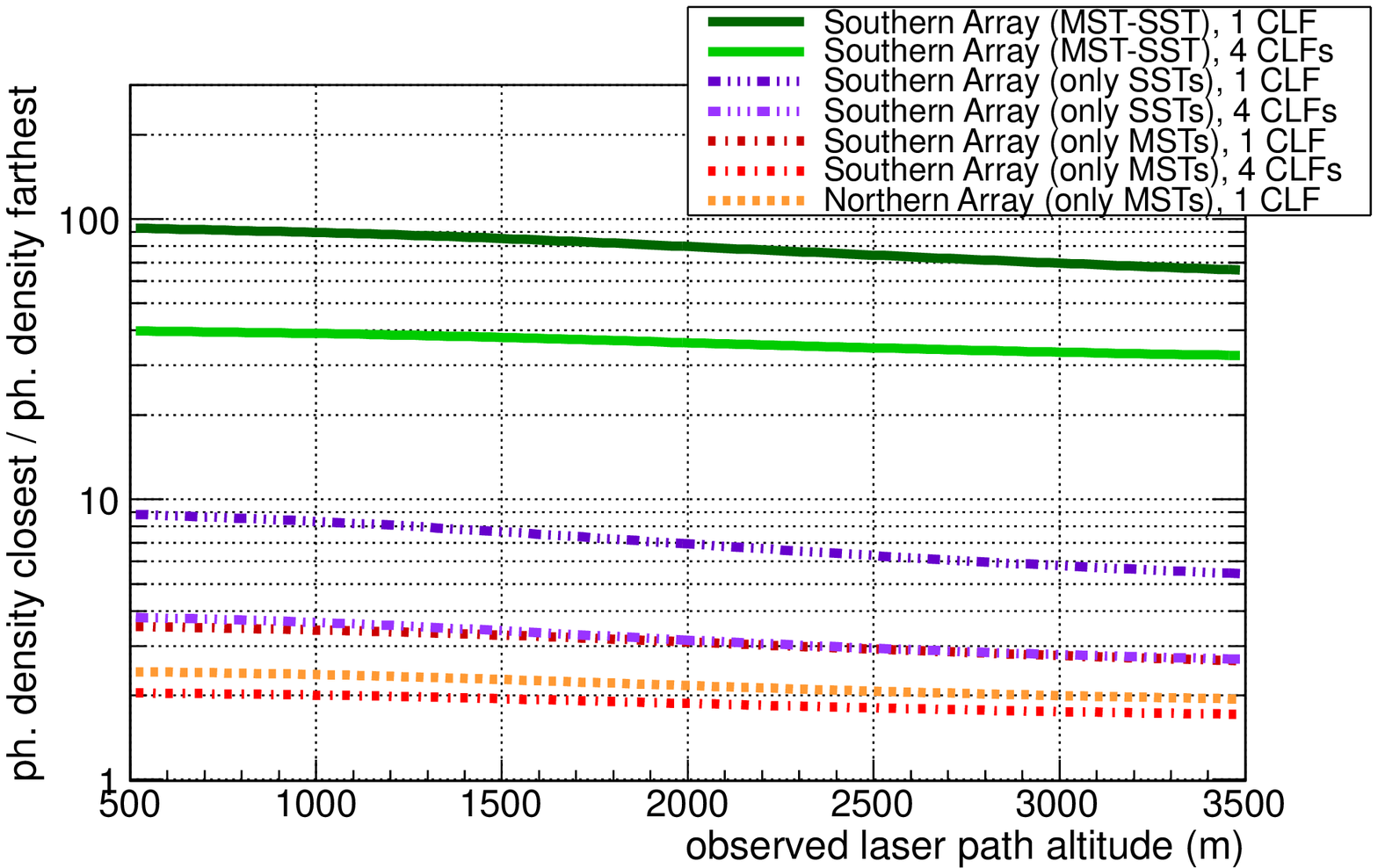}
\caption{The photon density (equation~\protect\ref{eq:Ipix:simp3}, divided by equation~\protect\ref{eq:dtfov}) from the closest telescope, divided by the photon density of the farthest telescope, 
for various observation altitudes and telescope types: The full green lines correspond to the closest medium-size-telescope, compared to the farthest small-size-telescope, 
the dashed-triple-dotted violet lines to the small-size-telescope only, and the dashed-dotted red lines to medium-size-telescopes only. 
The light green and light violet lines correspond to the same situation for the 
case of 4 central laser facilities, installed at each side of the array. 
The orange dotted line correspond to the Northern array, considering only medium-size-telescopes. \label{fig:rhodiff}}
\end{figure}

A possibility to reduce the differences in signal amplitudes consists in installing four CLFs, one at each side of the array, reducing these by another 
factor of about 2. In the best case (4 CLFs, observation height of 1~km), the signals differ then by maximally a factor 20. 
Another possibility to overcome 
this problem consists of a CLF designed with tunable laser power, such that linearity scans can be made, in which the usable intensity regimes of the different parts of the array
overlap.
%
%
In any case, the power of the CLF 
can be adjusted such that the signal received by the farthest small-size-telescope lies at a level of about 10 times the light of night sky, i.e. 15 photo-electrons during 
the roughly 100~ns that the laser pulse sweeps across the pixel, while the closest medium-size-telescope will then observe about 300 photo-electrons, still low enough 
not to enter saturation. 

Figure~\ref{fig:rhodiff} shows maximal ratios of photon density, i.e. the number of received photons, divided by the pulse width of one pixel. Here, higher laser observation 
altitudes are preferred since yielding small maximal differences. As in the previous case, the introduction of 4 CLFs reduces these differences by roughly a factor 2. 
If only telescopes of a same type are compared, the photon density of the farthest telescope w.r.t. the closest one results to lie within a factor 2 to 4, a value which 
can be handled easily, given the previous findings about the relative light collection efficiencies. 
%
%
%

\section{The calibration procedure}

The actual calibration procedure will consist of three steps:\\

1. Starting from the central pixel and moving slowly upward and downward the camera, 
the pulse of each pixel has to be found at the correct memory depth. Depending on the observed laser beam width, it may be necessary to 
include neighboring pixels' contents at the right and left side of the central line, until almost the entire beam width is covered.
The signal extraction procedure is the same as that used for ordinary $\gamma$-ray observations, including a correct subtraction of the 
baseline, pulse integration and the application of the flat-fielding correction factors $c_\mathrm{calib}$. As may be the case, a bandwidth correction factor 
$\epsilon_\mathrm{BW}$ has to be applied, 
which takes into account the bandwidth-limited AC-coupling of lower-frequency laser beam pulses, compared with the faster Cherenkov 
light pulses. Another correction factor $\epsilon_\mathrm{FOV}$ for the signal loss due to limited azimuthal coverage may be applied, 
especially when moving further outwards from the center of the camera, due to aberration effects. Starting from the integrated signal $S$, 
the flat-fielding corrected photo-electron content of the $i^\mathrm{th}$ pixel row is obtained as:

\begin{equation}
 I^{\,i} =  \left( \sum_{j=-n}^{+n} S^{ij} \cdot \epsilon_\mathrm{BW}^{ij} \cdot c_\mathrm{calib}^{ij} \right) \cdot \epsilon_\mathrm{FOV}^i \quad,
\end{equation}
where the summation goes over the $n$ pixels at both sides of the central line needed to integrate the full laser beam width.\\

2. The integrated pixel rows are then averaged and converted to an average number of photo-electrons per pixel row, by using the camera average photo-collection
 efficiency at the given laser wavelength. This procedure makes only sense if the calibration factors $c_\mathrm{calib}^{ij}$ from step 1 have been obtained by an external 
light source and therefore already reflect the differences in light-collection efficiency of the individual pixels.
In case of a hexagonal camera structure, an even amount of rows needs to be added, and a packing correction factor $\epsilon_\mathrm{pack}<1$ applied, 
which reflects the fact that two rows cover less than twice the pixel size in extension:

\begin{equation}
I_{pix} = \frac{\left( \sum_{i=-m}^{+m} I^{\,i} \right) \cdot \epsilon_\mathrm{pack} }{N} \quad,
\end{equation}
where the summation goes over the used $N$ pixel rows. \\

3. The result is then compared to the predicted average light illumination of one pixel, obtained from equation~\ref{eq:Ipix:simp3}.
Differences larger than the assumed statistical and systematic uncertainties will have to be attributed to an incomplete understanding of the 
term $A_\mathrm{eff,tel}$, which contains the effective mirror area $A_\mathrm{mirror}$, the average mirror reflectivity $\eta_\mathrm{mirror}$, 
light losses due to the protecting plexiglas of the camera and blind areas between pixels $\eta_\mathrm{loss}$ and the average photon detection 
efficiency $\eta_\mathrm{pde}$, thus:

\begin{equation}
A_\mathrm{eff,tel} = A_\mathrm{mirror} \cdot \eta_\mathrm{mirror} \cdot \eta_\mathrm{loss} \cdot \eta_\mathrm{pde} \quad .
\end{equation}

\section{Achievable precision}

The statistical precision of this calibration procedure depends mainly on the number of photo-electrons accumulated in those pixels which are used 
for the signal extraction. As we have seen in the previous section, a factor of at least 20 between signals from different telescopes must be assumed. 
In order to ensure that a Gaussian approximation of the photo-electron statistics is precise enough, 
the laser should be adjusted such that a pixel of that telescope which receives the faintest light pulse, receives at least 30--50 photo-electrons. 
On the other side of the array, a camera pixel will then receive on average at least 600-1000 photo-electrons.
If 10 rows can be used until the signal ranges 
out of the recorded memory depth, statistical uncertainties of 5--6\% for the furthest telescope and around 1\% for the closest telescope 
can be achieved for one laser shot. About hundred shots are needed to reduce the statistical error to below 1\% for all telescopes.
Additionally, the individual pixel calibration factors $c_\mathrm{calib}$ 
will have an uncertainty of maximally 5\% each, adding together another 1.5\% uncertainty. We will assume 2\% statistical uncertainty 
for here on. \\

\noindent
The following parameters can limit the achievable systematic precision of the CLF calibration procedure:

\begin{description}
\item[The laser output power $I_0$:\xspace] Commercially available standard lasers show an energy stability of $<2\%$, commercial Joule meters about 4\%. This value can be improved by the use of more than one independent (pyroelectric and photo-diode) probes to values of $<2\%$ (see e.g.~\cite{auger2011}).
\item[The FOV overlap correcion $\epsilon_\mathrm{FOV}$:\xspace] The precision of this value depends directly on the laser beam divergence and the precision with which 
the beam can maintain its direction. Commercial lasers can achieve 0.5~mrad beam divergences which result in a maximum beam width of 0.06$^\circ$.
Within this value, more than 90\% of the beam power will be found which ultimately fits into one pixel and can make the 
laser beam appear as one line through the camera, if the telescopes are focused to the correct distance and aberration effects are neglected. 
This is probably the case for Davies-Cotton mirrors over the major part of the FOV of the camera. In case of parabolic mirrors, an inner part 
of the laser image will have to be selected. 
\item[The beam direction:\xspace] Typically, beam directions can be maintained to within 0.04$^\circ$ degrees from vertical~\cite{auger2011}. Within these limits, an error
of always less than 1\% is introduced between the amount of light received by the closest and the farthest telescope to the viewed part of the laser beam.
\item[The atmospheric temperature and pressure:\xspace] The Rayleigh scattering cross section depends critically on the air density, which can be 
 measured from temperature and pressure at height $h$, assuming that the air behaves like an ideal gas. Commercial 
radio sondes provide accuracies of 0.5~K and 1~hPa. 
This translates into a relative uncertainty of the term $P(h)/P_s \cdot T_s/T(h)$ of about 0.2\%. Without radio sonde measurements, 
the Global Data Assimilation System (GDAS) could be used, as shown in~\cite{gdasauger:2012}, where accuracies of about 1.5\% for $T_s/T(h)$ and 0.2\% for $P(h)/P_s$ have 
been obtained. If at the site a Raman Lidar is operated, the systematic uncertainty of the molecular density can be reduced to negligible quantities~\cite{Wandinger:2005}. 
We can further assume that a professional meteorological station will be operated at the site, yielding pressure and temperature at the telescope altitude with 
accuracies of better than 0.1\%. The translation of these measurements to an altitude of 2000~m above ground should be precise to better than 1\% at night, when no 
temperature inversion is expected and dry conditions and no clouds are assumed (see e.g.~\cite{rolland}).
%
  
Without any knowledge about the local atmosphere, accuracies of not better than 20\,\% can be obtained~\cite{tomasi}.
%
%
\item[The vertical optical depth:\xspace] At UV wavelengths and clear astronomic nights, the optical depth is dominated by 
molecular scattering, rather than aerosols (see e.g.~\cite{patat}). 
As shown in the previous paragraph, molecular extinction can be modelled to the same precision, as pressure and temperature profiles from the scattering 
point to the telescope are understood, i.e. to about 1\%, if adequate instrumentation is available. Transmittances of 0.65 to 0.82 are then expected, 
depending on the respective distances. 
Extinction of UV light due to aerosols should never exceed 5\% for any astronomic site during clear nights, except maybe for the Namibian candidate sites for the CTA. 
%
%
The AUGER collaboration~(fig. 9 of~\cite{auger2013}) shows typical systematic uncertainties to the vertical aerosol optical depth at 2~km of 
the order of 0.005 for clear nights, for an average optical depth of about 0.02. 
\cite{auger2011} shows that a typical scatter from the day-by-day measurements of the vertical aerosol optical depth amounts 
to about 30\% for the clearest nights. Both translate to an uncertainty of less than 1\% for the aerosol transmission from 2~km to ground. 
The transmittances should be well under control if the CLF is operated together with additional instrumentation, like a Raman LIDAR, 
otherwise an uncertainty of about 2\% must be assumed, except for the case that the CTA is installed at a Namibian site, where aerosols are less well characterized. 
In that case, an uncertainty of 5\% should be conservatively used, at least until extensive aerosol characterization campaigns are led there.
\item[The aerosol volume scattering cross section:\xspace] This parameter is cited as the main limiting factor by the AUGER and TA 
experiments~\cite{auger2006, augerabsolute2,ta2008}, reaching precisions better than 10\% only with the use of an accompanying LIDAR.  
Systematic monitoring of the vertical aerosol optical depth
above ground made by the AUGER experiment~\cite{auger2013} shows that practically aerosol-free nights could be achieved regularly, with an average
aerosol vertical optical depth of about 0.025 from ground to 2~km altitude. This value then increases more or less linearly to $\tau_\mathrm{aer} \approx 0.047$ 
from ground to 4~km. Very clear nights however show much lower optical depths, namely only $\tau_\mathrm{aer} \approx 0.007$ from ground to 4~km altitude. 
These numbers translate roughly to $\beta_\mathrm{aer} \approx 1.75\cdot 10^{-6}\, \mathrm{m}^{-1}$ for clear nights and 
$\beta_\mathrm{aer} \approx 1.0\cdot 10^{-5}\, \mathrm{m}^{-1}$ for the average case above 2~km a.g.l. The former yields the relative importance of 
aerosol scattering w.r.t. Rayleigh scattering as shown in figure~\ref{fig:epsaer}, which is hence well under control if UV wavelengths are used for the CLF laser.
``Clear'' nights introduce an uncertainty of the order of 1--2\% to an absolute calibration 
with a CLF, whereas ``typical'' nights may degrade that uncertainty to the order of 5\%, even if the aerosol optical depth is known, due to unknowns in the phase function. 
Very clear nights can serve to reduce the contribution of aerosol scattering to less than one percent only. 
\item[The bandwidth of the signal amplification chain:\xspace]  As seen in section~\ref{sec:geom}, the recorded signal from the laser beam in one single pixel can 
be as large as 120~ns. The amplification chain must be able to amplify such a large signal in exactly the same way as it would do for nano-second pulses, typical for 
Cherenkov showers. Especially high-pass filters in the signal amplification chain must have a cut-off frequencies well below 10~MHz, 
otherwise correction factors $\epsilon_\mathrm{BW}$ have to applied. 
As the signal durations will be different for the different types of telescopes, the inter-calibration between the telescope types will be affected.
\item[Background light leakage:\xspace] To quantify the effect of the background light leakage, 
precise  knowlegde of the typical light-of-night sky at the CTA site is required.
Especially large mirror telescopes and large pixel sizes will be affected. With the numbers given in table~\ref{tab:tellist}, the medium-size telescope 
would receive most of the background light. We estimate its contribution, 
starting from typical background light measurements with the MAGIC telescopes (0.13 photo-electrons per pixel per nano-second (see e.g.~\cite{bartko}), 
and obtain approximately 0.19 and 0.03 photo-electrons per nano-second background light contribution for the medium-size and the small-size telescopes, respectively. 
Since the ratio in photon densities received by one (medium-size) telescope 
and another (small-size) telescope is a maximally a factor 20 (see figure~\ref{fig:epsdiff}), if 4 CLFs are used in the South, and readout saturation starting 
from mean pulse levels of 700 photo-electrons is assumed, then a pixel of the farthest small-size telescope would receive about 35 photo-electrons, during about 100~ns. 
In that pixel, the contribution of background light amounts then to $\sim$10\% of the CLF signal, if the beam width spreads over only one pixel. 
Otherwise, the relative contribution is higher. 
A similar calculation only for the medium-size telescope yields a 3.5\% contribution of background light.
Appropriate baseline subtraction will reduce this number again, 
however stars and other sources of background light may spoil the efforts again. 
Generally, it is preferable to have large (not yet saturating) signal pulses which minimize the relative importance of the background light.
\item[The residual polarization of the laser beam:\xspace] We assume here a CLF able to emit light pulses with less than 2\% polarization, 
and neglect the azimuthal dependence of the Rayleigh scattering coefficient. Values in this range are possible if a suitable depolarizer is used (see e.g.~\cite{ta2008}). 
The precision can be increased if the whole laser can be rotated by 90$^\circ$ and the measurements repeated.
\item[Spectral contamination of other harmonics:\xspace]  Nowadays available commercial lasers yield spectral purities of better than 99\%, if a Nd:YAG laser is used. 
In case of a nitrogen laser, this problem does not appear.
\item[Simplifications of equations:\xspace] The simplifications in equations~\ref{eq:Ipix:simp1} and~\ref{eq:Ipix:simp3} introduce systematic errors of maximally 2\% and 
can eventually be reduced by appropriate MC simulations of the scattered laser light. 
\end{description}

\noindent
Table~\ref{tab:systerr} shows the magnitude of all systematic uncertainties and which type of calibration they affect. 
All numbers reflect best guess estimates which will probably be reduced with time and experience and additional hardware 
to measure each contribution separately. Including statistical uncertainties, an absolute calibration of about 4--5\% uncertainty 
seems possible, if the camera hardware is adapted to the requirements of a CLF. Additionally, each monitoring point will fluctuate the same amount. 
Note especially that the numbers given for the inter-telescope 
and inter-telescope type calibration reflect the uncertainty for each calibration run and can probably be reduced by applying 
the calibration procedure throughout different nights. 

\begin{table}[htp]
\centering
\begin{small}
\begin{tabular}{l|l|l|l}
Source uncertainty & size & type of calibration affected & remarks  \\
\hline
Statistical/$c_\mathrm{cal}$ & 1--2\%   &  ABS, I-TEL  &        \rule{0mm}{3mm}  \\
Laser power        & 2\%       & ABS, TIME &         \rule{0mm}{3mm}\\
$\epsilon_\mathrm{FOV}$ &  $<$10\% & ABS, I-TEL, I-TYPE & depends on telescope focus  \rule{0mm}{3mm} \\
beam direction    & 1\%        & ABS, TIME, I-TEL, I-TYPE &       \rule{0mm}{3mm}\\
$P(s), T(s)$      & 1\%   & ABS, TIME & depends on atmospheric  \rule{0mm}{3mm}\\
                  &       &           &  monitoring equipment   \\
$\tau_\mathrm{mol+aer}$ & 1--5\%  & ABS, TIME, I-TEL & clearest nights,   \rule{0mm}{3mm}\\
                     &         &                & Lidar required \\
$\beta_\mathrm{aer}$ & 1--5\% & ABS, TIME, I-TEL & depends on atmospheric \rule{0mm}{3mm}\\
                    &         &                 & monitoring equipment  \\
                    &         &                 & 1\% I-TEL for Southern array only \\
$\epsilon_\mathrm{BW}$ &  0--10\%  & ABS, I-TYPE & depends on  \rule{0mm}{3mm}\\
                   &          &          & electronic coupling \\
BG light  & 2--10\% & ABS, TIME, I-TEL, I-TYPE & depends on FOV: lower value \rule{0mm}{3mm}\\
          &        &                      & for carefully selected pointings\\
beam polarization  & $<$2\% & ABS, I-TEL &  reduced by turning the laser  \rule{0mm}{3mm}\\
spectral contam. & $<$1\% & ABS   & only for Nd:YAG lasers  \rule{0mm}{3mm}\\
Rayleigh scatt. eq.~\ref{eq:Rayleigh} & 0.5\%  & ABS, I-TEL    & I-TEL only for Southern array   \rule{0mm}{3mm}\\
                                      &        &               & negligible for Northern array   \\
\hline
\end{tabular}
\end{small}
\caption{List of sources of systematic errors for the CLF calibration types: absolute calibration of the whole array (ABS), 
time evolution of $A_\mathrm{eff,tel}$ for each individual telescope (TIME), inter-telescope calibration (I-TEL) and 
calibration of the three different telescope types w.r.t. each other (I-TYPE).  \label{tab:systerr}}
\end{table}

\section{Hardware Considerations}

Considering the previous findings, a CLF will do its work only if 
several hardware requirements are met on the side of the CTA telescopes:

\begin{description}
\item[Focuses down to 3~km distance:\xspace] Each telescope should have the possibility to focus to a distance of about 3~km or lower. 
Otherwise, the light beam cannot be contained in only one row of pixels, and the subtraction of backgrounds becomes much more unprecise.
\item[Different readout depths:\xspace] Each pixel along the viewed laser path receives the signal at different times. This requires a different readout depth 
for each channel along the signal axis, or the possibility to adjust the readout time window beforehand. Especially the central pixels must be fully contained 
inside the signal digitization window, although the outmost pixels, where the laser pulse image enters into the camera, will launch the majority trigger. In the case 
of an external trigger, this requirement becomes less strict.
\item[Correct amplification of long pulses:\xspace] The photo-multiplier front-end electronics needs to be able to amplify and electronically transmit longer signal pulses
in the same way as the short Cherenkov light pulses. Especially AC-couplings along the electronic signal transmission chain must leave enough bandwidth for pulses as 
long as 100~ns, in order not to be distorted or even filtered.
\item[Digitization of long pulses:\xspace] A signal digitization mode must be available which allows integration of the entire pulse from CLF runs, which can last as long as 100~ns.
\item[External or differential trigger:\xspace] Unless the readout windows can be configured on micro-seconds time-scales, an external trigger needs to be installed for each telescope individually 
to trigger the CLF readout such that the central pixels'  
signals get correctly centered in time. Alternatively, a differential trigger could trigger the readout of each pixel (or cluster of pixels) at the 
correct time delay w.r.t. to the previously hit pixel (or pixel cluster).
\end{description}
\noindent

\begin{figure}[htp]
\centering
\includegraphics[width=0.495\linewidth]{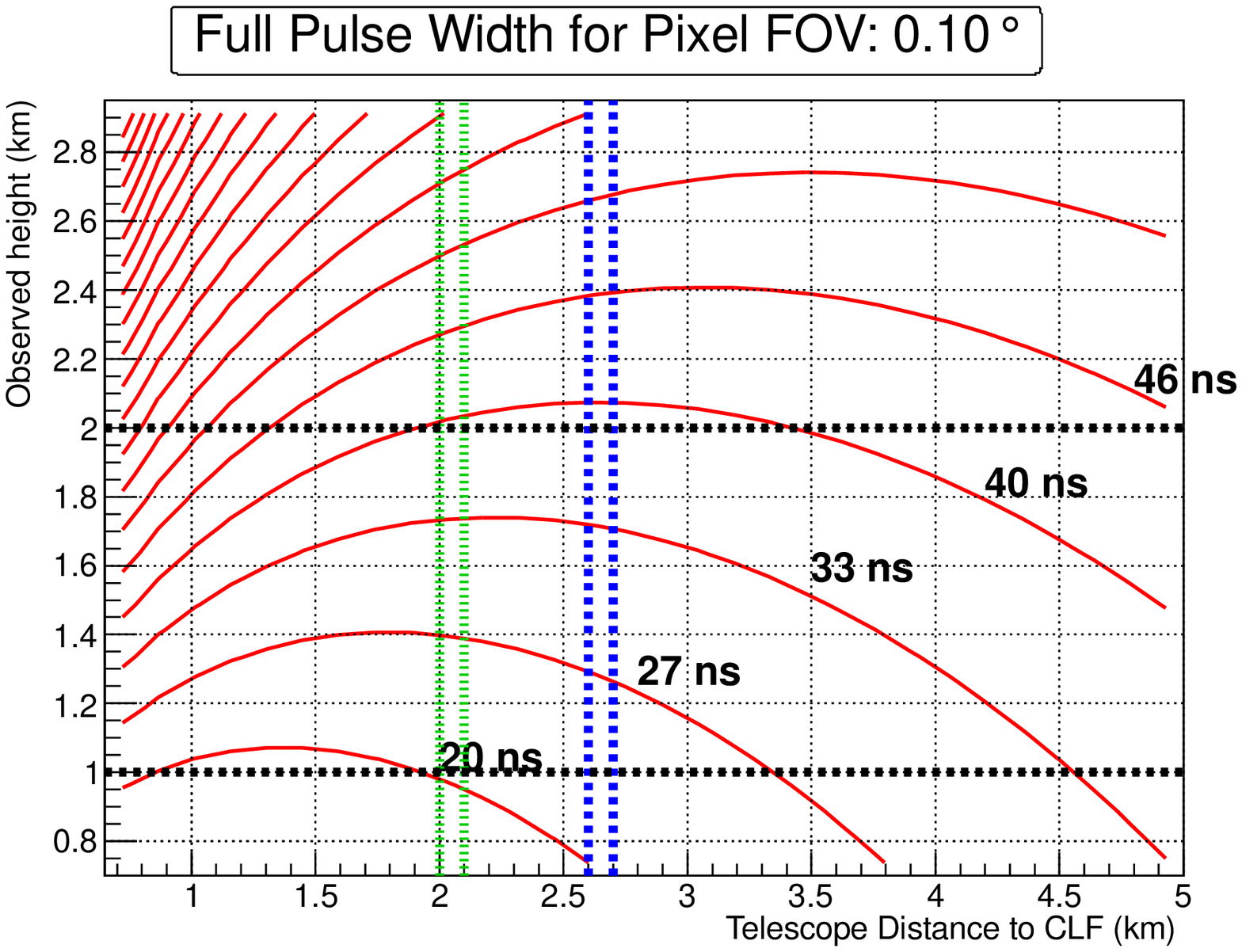}
\includegraphics[width=0.495\linewidth]{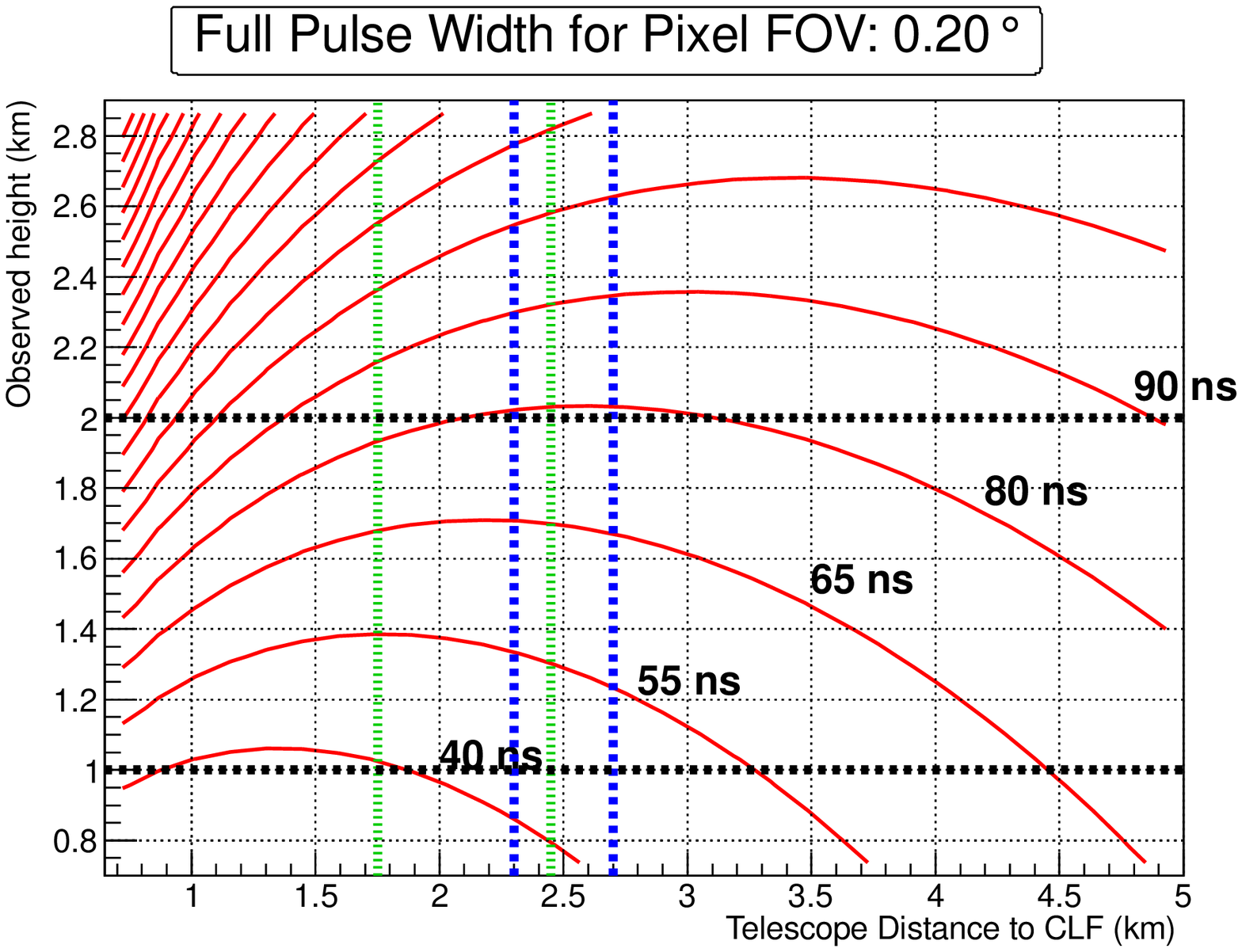}
\includegraphics[width=0.495\linewidth]{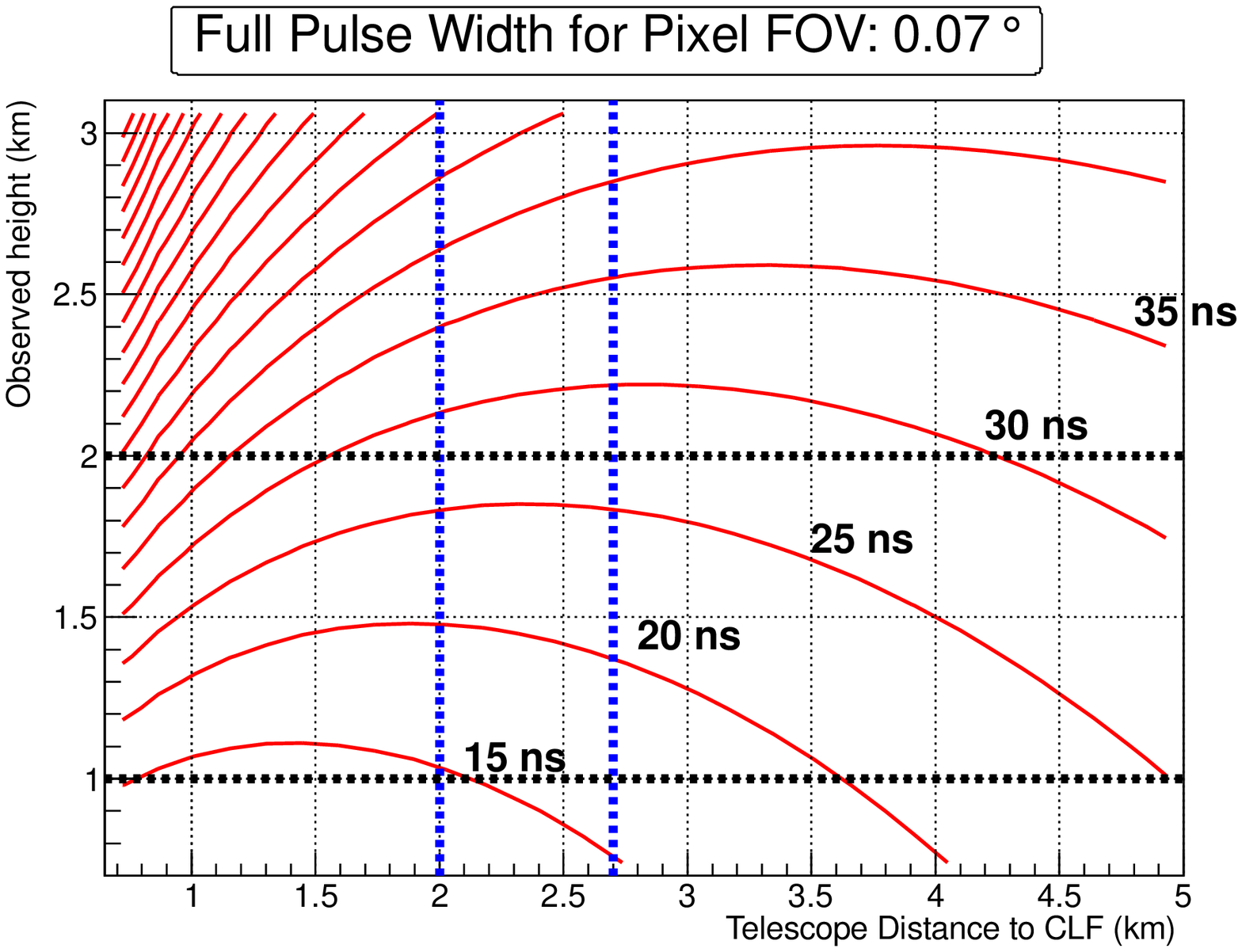}
\includegraphics[width=0.495\linewidth]{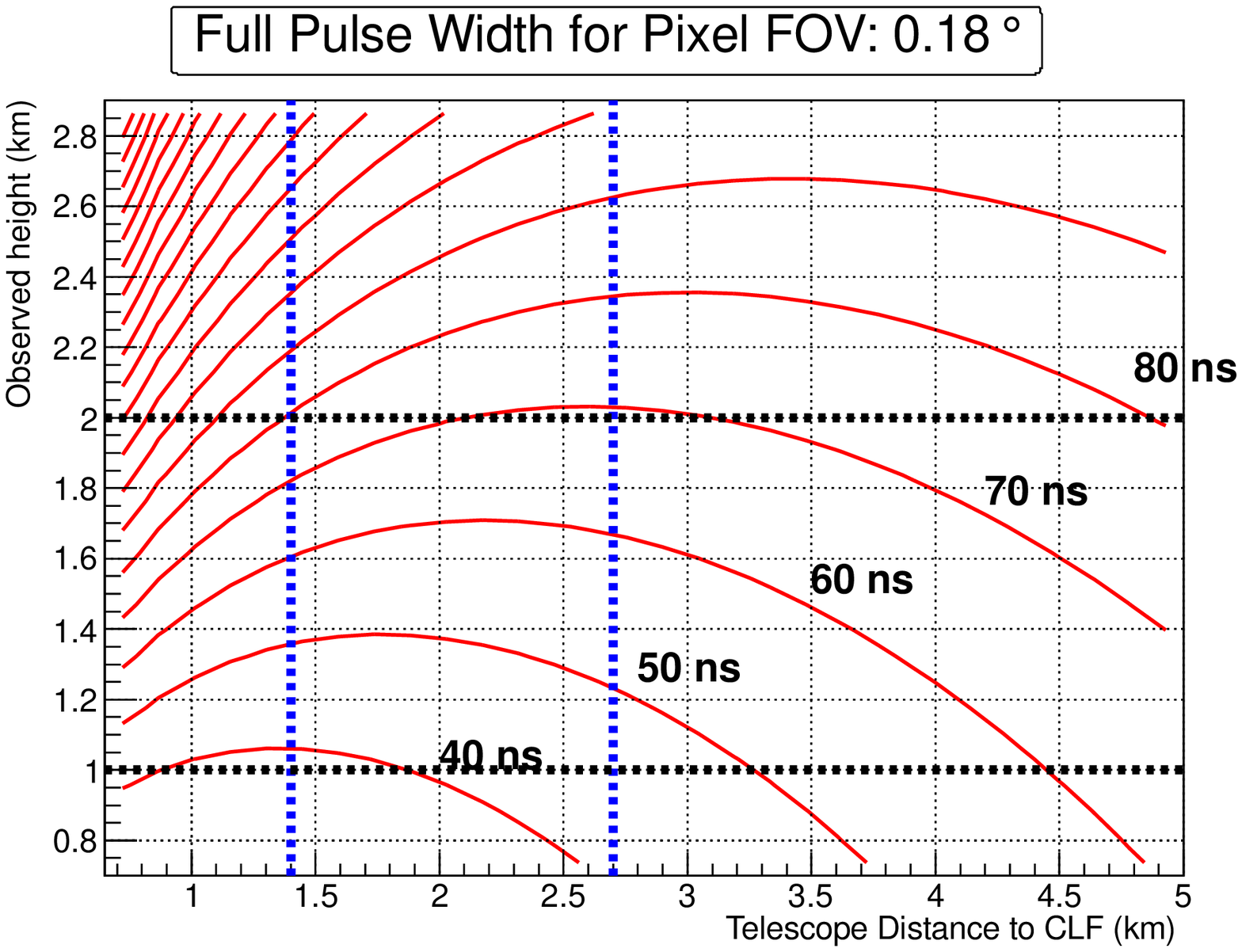}
\includegraphics[width=0.495\linewidth]{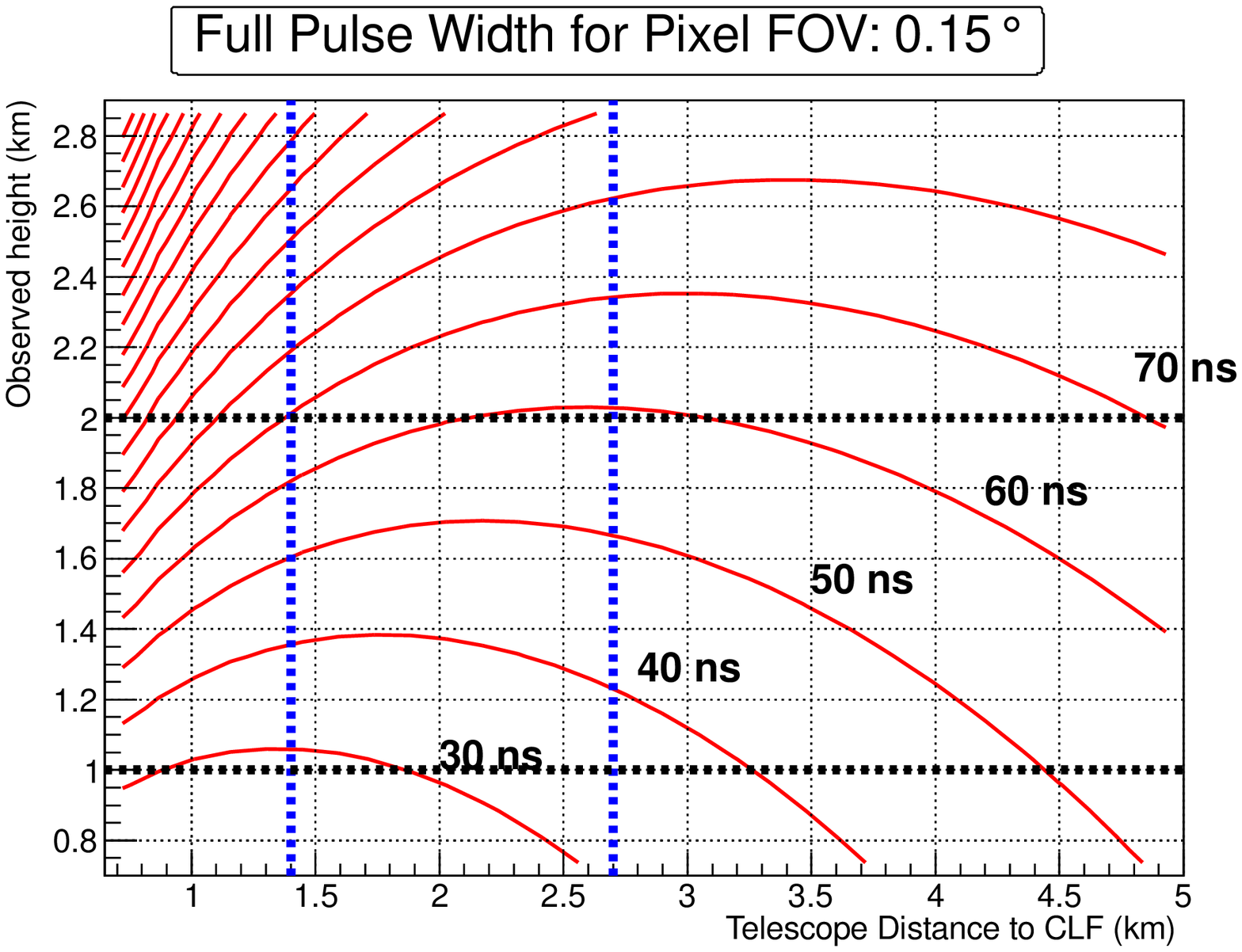}
\includegraphics[width=0.495\linewidth]{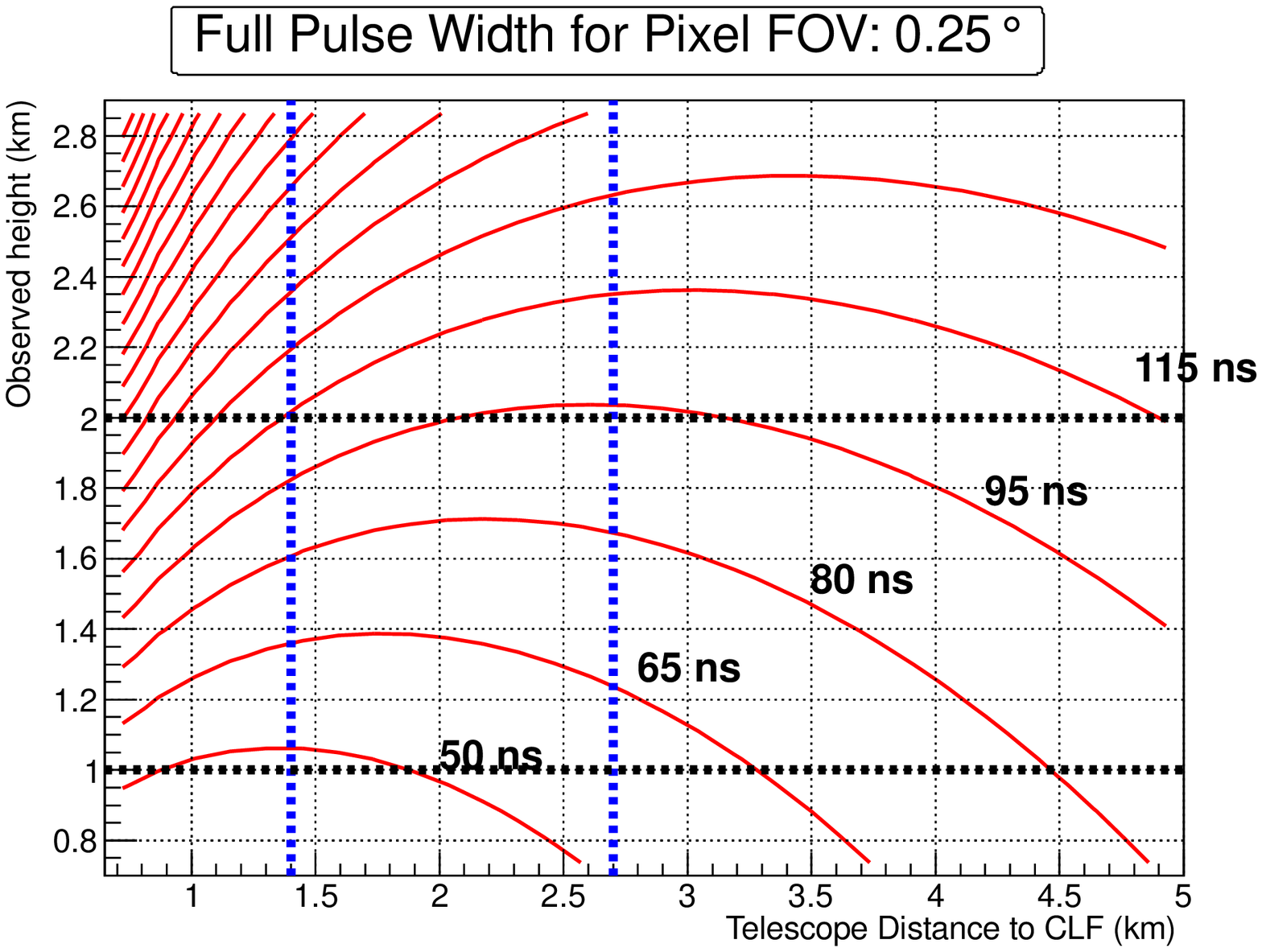}
\caption{Contours of eq.~\protect\ref{eq:dtfov} plotted for typical pixel sizes, corresponding to the current design parameters of the 
large-size telescope \protect\cite{teshima2013} (top left), the medium-size telescope \protect\cite{ctaconcept} (top right), the mid-size Schwarzschild-Couder 
telescope~\protect\cite{vasiliev2013} (center left), the small-size-2mirror telescope, \cite{pareschi2013} (center right), 
the small-size-2mirror telescope \protect\cite{zech2013} (bottom left) and the small-size 1mirror telescope~\protect\cite{moderski2013} (bottom right), as a function of 
the observed height of the laser path and the telescope distance from the CLF.
The blue dashed lines indicate the area of the Southern array~\protect\cite{ctaconcept}, covered by one out of 4 CLFs, the green dotted lines show the same for the Northern array. 
The horizontal black lines indicate the optimum case (2~km observation height), and an absolute minimum requirement (1~km observation height).
 \label{fig:tpix} }
\end{figure}

Based on the points from the previous list, minimum requirements for the telescopes cameras can be formulated.
Figure~\ref{fig:tpix} shows the pulse widths seen for different camera pixel sizes, depending on the telescope distance from the CLF and the observed 
laser path height $h$. Once the optics are fixed and the final extension of the array known, a minimum pulse width registration capability can be derived for 
an optimum case ($h = 2$~km) and an absolute minimum case ($h = 1$~km). The results are shown in tables~\ref{tab:optimum} and~\ref{tab:minimum}, respectively.
One can see immediately that the obtained values are always larger than the 15–20~ns per pixel time window, required to achieve optimal Cherenkov light collection~\cite{ctamc}. 
The situation is most critical for the small-size telescope with one Davies-Cotton mirror, 
due to the larger pixel FOV and the physical extent of the array in the South.  
Using several CLFs, e.g. located at each side of the array, could help to reduce the minimum requirements, while the optimum case becomes almost independent of the number of CLFs.

\begin{table}[htp]
\centering
\begin{tabular}{|l|c|c|c|}
\hline
    & \multicolumn{3}{|c|}{Minimum pulse width registration capability} \\
\hline
    & CTA-North  & CTA-South & CTA-South \\
    &            & (4 CLFs)  & (1 CLF)   \\
\hline
Large-Size-Telescope                 &  45~ns      &   45~ns   &  45~ns  \\
\cite{teshima2013} &             &            &    \\
Medium-Size-Telescope                  &  85~ns     &   85~ns   &  90~ns  \\
\cite{ctaconcept}  &             &            &    \\
Mid-Size Schwarzschild-Couder-Tel.   &             &   30~ns   &  35~ns  \\
\cite{vasiliev2013} &            &            &    \\
Small-Size-Telescope-2Mirror     &             &   85~ns   &  90~ns  \\
\cite{pareschi2013} &            &            &    \\
Small-Size-Telescope-2Mirror     &            &   70~ns   &  75~ns  \\
\cite{zech2013}     &            &             &    \\
Small-Size-Telescope-1Mirror    &            &   120~ns  & 125~ns  \\
\cite{moderski2013} &            &           &   \\
\hline
\end{tabular}
\caption{Requirements for an optimum case (considering $h = 2$~km) for the current design parameters of the different telescopes.  
\label{tab:optimum} }
\end{table}

\begin{table}[htp]
\centering
\begin{tabular}{|l|c|c|c|}
\hline
    & \multicolumn{3}{|c|}{Minimum pulse width registration capability} \\
\hline
    & CTA-North  & CTA-South & CTA-South \\
    &            & (4 CLFs)  & (1 CLF)   \\
\hline
Large-Size-Telescope                 &  25~ns      &   30~ns   &  30~ns  \\
\cite{teshima2013} &             &           &    \\
Medium-Size-Telescope                  &  50~ns     &   55~ns   &  60~ns  \\
\cite{ctaconcept}  &             &            &    \\
Mid-Size-Schwarzschild-Couder-Tel.    &             &   20~ns   &  25~ns  \\
\cite{vasiliev2013} &            &            &    \\
Small-Size-Telescope-2Mirror     &             &   50~ns   &  60~ns  \\
\cite{pareschi2013} &            &            &    \\
Small-Size-Telescope-2Mirror     &         &   40~ns   &  50~ns  \\
\cite{zech2013}     &            &            &    \\
Small-Size-Telescope-1Mirror    &             &   65~ns   &  85~ns  \\
\cite{moderski2013} &            &            &    \\
\hline
\end{tabular}
\caption{Minimum requirements (considering $h = 1$~km) for the current design parameters of the different telescopes.  
\label{tab:minimum} }
\end{table}

This allows the derivation of an absolute minimum requirement for the correctly amplified and registered pulse width of 20 to 65~ns, depending on the telescope type, for 
the Southern array, and 25 to 50~ns for the Northern array, respectively. The same numbers apply to the triggerable time delay between 
two pixels.


Finally, combined systematic uncertainties for different hardware solution scenarios are shown in table~\ref{tab:systerrtot}, 
for the extended Southern and the smaller Northern array, and separately for the absolute calibration of the array, 
inter-telescope and inter-telescope-type calibration and for the sensitivity monitoring of each individual telescope. 
One can see that adapted camera hardware is more critical for the Southern array, due to the installation of the small-size 
telescopes with large FOV, and the wider spread of telescopes and hence overall larger differences in distance to the CLF.
The atmospheric calibration with a Lidar seems not so stringent, if clear nights are chosen for CLF calibration runs 
and the local atmosphere at the observatory is monitored via weather stations.

\begin{table}[htp]
\centering
\begin{tabular}{|l|c|c|c|c|c|c|}
\hline
Calibration  & adapted     &  No     &  No       &  No    & Low     & Wave- \\
Type         & telescopes, & focus   & adapted   & Lidar  & alti-  & lengths \\
             & Lidar,      & to      & band-     &        & tude   & $>$400~nm  \\
             & selected    & 3~km    & width     &        & site,  &             \\
             & pointings,  &         &           &        & high   &             \\
             & small       &         &           &        & aerosol &             \\
             & aerosol     &         &           &        & content &             \\
             & content     &         &           &        &         &             \\
\hline
\multicolumn{6}{c}{Southern Array} \\
\hline
ABS          &   4\%                &  6\%    & 11\%      &   5\%  & 8\% & 5\% \\ 
I-TEL        &   4\%                & 11\%    &  4--8\%   &   4\%  & 7\% & 5\%\\
I-TYPE       &   3\%                &  6\%    &  10\%     &   3\%  & 5\% & 3\% \\
TIME         &   4\%                &  4\%    &  4\%      &   4\%  & 8\% & 5\% \\
\hline
\multicolumn{6}{c}{Northern Array} \\
\hline
ABS          &   4\%                &  5\%    & 6--7\%    &  5\%  & 8\% & 5\% \\ 
I-TEL        &   3.5\%              &  8\%    & 4--5\%   &   4\%  & 6\% & 4\%\\
I-TYPE       &   3\%                &  5\%    &  6 \%     &  3\%  & 4\% & 3\% \\
TIME         &   4\%                &  4\%    &  4\%      &  4\%  & 8\% & 4\% \\
\hline
\end{tabular}
\caption{Combination systematic uncertainties for the CLF calibration types: absolute calibration of the whole array (ABS), 
time evolution of $A_\mathrm{eff,tel}$ for each individual telescope (TIME), inter-telescope calibration (I-TEL) and 
calibration of the three different telescope types w.r.t. each other (I-TYPE). The consequence of different levels of hardware adaptation
on the systematic uncertainty are shown. \label{tab:systerrtot}}
\end{table}

\section{Discussion}

Table~\ref{tab:systerrtot} summarizes the potential to calibrate the entire array, including all types of telescopes, 
in a fast and precise manner. Parts of this goal may also be achieved by other means, used for past and present IACTs. In this section, 
we will discuss how the precision and the impact of a CLF-based calibration scheme compares with these other techniques.

Calibration methods for individual telescopes with systematic uncertainties in the range of 10--15\% have been proposed in~\cite{frass,puhlhofer,lebohecatmosphere,karschnik}
 and are not further considered here since they are either too unprecise or require hardware unsuitable for an array of many telescopes, or demand too much 
downtime to calibrate the entire array. 





Elaborate methods comparing images recorded from local muons with those from simulations 
have been developed~\cite{rose,puhlhofer,leroy,goebel,bolz}, reaching a precision down to a few percent. 
To achieve this, shadowing of the mirrors by the camera or masts needs to be simulated thoroughly, 
and the effect of the different emission spectra 
(starting from 250~nm wavelength in the case of muons, while gamma-ray shower light is typically absorbed below 300~nm) needs to be corrected for~\cite{humensky}.
Muons lose only 3.5--4\% of the light in the range from 300 to 600~nm from the point of emission to the photo detector in the camera~\cite{bolz} and are hence 
much less affected by the corresponding atmospheric uncertainties, if compared to a CLF. It has to be clear however, that calibration methods 
using muons also require some hardware adaptation from the individual telescopes trigger. Otherwise, muons may be completely lost since the may be rejected by the 
multi-telescope trigger or a too high threshold. Furthermore, it is still not clear whether the small-size telescopes will be sufficiently sensitive to register useful, i.e. 
unbiased, muon images for calibration. Finally, a CLF can provide calibration at distinct wavelengths, while the Cherenkov light spectrum from muons cannot be changed.  
It further allows for light intensity scans, albeit over a limited intensity range for every telescope, but with overlapping regions between each. 


In the low-energy range, cross-calibration with sources observed by the (more precisely calibrated) 
FERMI satellite can yield a precision of ${}^{+5\%}_{-10\%}$~\cite{meyer,bastieri}. 
This will calibrate the energy reconstruction of the large-size and combinations of medium-size telescopes, while the small-size telescopes 
need to rely on other known spectral features at higher energies, such as the cut-off in the cosmic-ray electron spectrum, measured by PAMELA and AMS. 

Using cosmic ray images, and particularly the distributions of their image sizes and reconstructed shower impact points, 
\cite{hofmann} claims a precision of 1--2\% for inter-telescope calibration. As in the case of muons, the method seems superior in terms of precision 
and does not need any hardware adaptation, however information about the spectral sensitivity cannot be derived, and overall degradations of the array 
cannot be detected, especially if the atmosphere is not understood to the same level of precision. At the current understanding of hadronic cosmic ray 
showers, it seems very challenging to achieve an absolute calibration of the array to a level better than 10\% with that method. 

In summary, it seems useful to operate an CLF for combined and fast calibration of the array. 
Although other methods may result in a partially more precise 
calibration, these cannot provide inter-telescope, inter-telescope-type, absolute and spectral calibration at a same time. Since at the aimed 
performance of the CTA different calibration methods must be cross-checked with others, a CLF presents itself as a reasonable candidate.

\section{Conclusions}

This study shows that a Central Laser Facility can be a solution for a fast calibration of the CTA, 
if certain design requirements for the CTA cameras are met. 
These must provide either externally triggered individual readout windows for each pixel or a pixel-cluster trigger configurable 
in such a way that the laser beam image subsequently triggers small units of pixels on its path through the camera. 
Accordingly, the minimum accessible memory depth must be correspondingly large, at least 1~$\mu$s, and the readout 
must be able to reconstruct pulses of up to 120~ns length without distortion 
due to the internal AC-couplings of the readout and amplification chain. In this case, the laser facility can calibrate all telescopes at the same time, on a time 
scale which is mainly limited by the movement of the telescopes towards the individual pointing toward the laser beam. Such a calibration 
can be run every night, although very clear nights are favored to reduce systematic uncertainties due to the aerosol contribution to the scattering process 
and the subsequent extinction of the scattered light on its way to the telescope cameras. 
A CLF can then be used to monitor the sensitivity of each telescope, 
to calibrate the telescopes among each other as well as to achieve a relative calibration between the different telescope types, a task which 
is difficult to achieve by other means. 
Especially appealing is the possibility to use different laser wavelengths to calibrate the spectral sensitivity of the array. 
Finally, an absolute calibration with a precisions ranging from 4--11\% seems possible, depending on the adaptation of the telescope hardware 
and if additional atmospheric monitoring devices, such as a LIDAR, are operated together with the CLF.



\acknowledgments

The support from the MICINN through the National R+D+I, Spain, and 
the support of the Comissionat per a Universitats i Recerca del Departament d'Innovaci\'o, Universitats i Empresa, 
de la Generalitat de Catalunya and of the Cofund program of the Marie Curie Actions of the 7$^\mathrm{th}$ Framework Programme for Research
of the European Union are gratefully acknowledged. The author thanks also Garret Cotter, Aurelio Tonachini, Piero Vallania and Laura Valore for thorough revision 
of the manuscript.

\bibliography{biblio}
\bibliographystyle{JHEP}

\end{document}